\documentclass[showpacs,pre,twocolumn,numbers,showkeys,superscriptaddress]{revtex4-1}

\usepackage[T1]{fontenc}
\usepackage[utf8]{inputenc}
\usepackage[english]{babel}

\usepackage{listings}
\usepackage{bold-extra}
\usepackage{color}
\usepackage{amsmath,amsthm} 		% automatically loaded by amsart
\usepackage{amssymb} 			% additional math fonts
\usepackage[flushmargin,hang]{footmisc}
\usepackage{mathrsfs,fancyhdr,latexsym}
\usepackage{graphicx}
\usepackage[normalem]{ulem}

\newcommand{\e} {\text{e}}                  
                  
\newcommand{\eps} {\varepsilon}

\newcommand{\w}{\omega}

\newcommand{\ph}{\varphi}

\newcommand{\tht}{\theta}

\newcommand{\poinc}{Poincar\'e }

\newcommand{\crit}{_{\rm crit}}

\newcommand{\K}{K$^{+}$}

\begin{document}

\title{Qualitative and Quantitative Stability Analysis of Penta-rhythmic Circuits}

\author{Justus T.~C.~Schwabedal}
\affiliation{Neuroscience Institute, Georgia State University, 100 Piedmont Ave SE, Atlanta, Georgia 30303, USA}
\affiliation{Max Planck Institute for the Physics of Complex Systems, N\"othnitzer Str.~38, 01187 Dresden, Germany}

\author{Drake E.~Knapper}
\affiliation{Department of Physics, Georgia State University, 25 Park Place, Atlanta, Georgia 30303, USA}
\affiliation{Neuroscience Institute, Georgia State University, 100 Piedmont Ave SE, Atlanta, Georgia 30303, USA}

\author{Andrey L.~Shilnikov}
\affiliation{Neuroscience Institute, Georgia State University, 100 Piedmont Ave SE, Atlanta, Georgia 30303, USA}
\affiliation{Department of Mathematics and Statistics, Georgia State University, 33 Gilmer Street SE, Atlanta, Georgia 30303, USA} 
\affiliation{Institute for Information Technology, Mathematics and Mechanics, Lobachevsky State University of Nizhni Novgorod, Nizhni Novgorod, 603950,  Russia} 
 
%\pacs{
%87.19.ln oscillations and resonance in biological systems      ,
%05.45.Xt synchronization, coupled oscillators                  ,
%05.70.Ln nonequilibrium, irreversible processes                ,
%05.10.Gg stochastic analysis methods                           ,
%05.10.-a computational methods in statistical physics          ,
%87.18.Tt noise in biological systems                           ,
%87.10.Mn stochastic biological physics                         ,
%}

\begin{abstract}
Inhibitory circuits of relaxation oscillators are often-used models for the
dynamics of biological networks.  We present a qualitative and quantitative
stability analysis of such a circuit constituted by three reciprocally coupled
oscillators of a Fitzhugh-Nagumo type as nodes.  Depending on inhibitory
strengths, and parameters of individual oscillators, the circuit exhibits
polyrhythmicity of up to five simultaneously stable rhythms.  With methods of
bifurcation analysis and phase reduction, we investigate qualitative changes in
stability of these circuit rhythms for a wide range of parameters.
Furthermore, we quantify how robustly rhythms are maintained under random
perturbations by monitoring phase diffusion in the circuit.  Our findings allow
us to describe how circuit dynamics relate to dynamics of individual nodes.  We
also find that quantitative and qualitative stability of polyrhythmicity do not
always align.  
\end{abstract}

\maketitle

\section{Introduction}

Relaxation oscillators have become a base for constituting coupled
systems that generate a variety of self-sustained rhythms.  These types of
oscillators are used to model heart beats \cite{vdPol1928}, cellular membrane
dynamics \cite{Fitzhugh1961}, electrical and mechanical systems
\cite{Mandelstam1969,Andronov1987}, harmful algae bloom \cite{Franks1997},
regulatory genetic networks \cite{Hasty2001, Guantes2006}, and the population
dynamics of pest cycles of forests \cite{Brons2010a}.

Simple relaxation oscillations are formed by two essential variables: one
\textit{activity variable} assumes active and inactive states, and one
\textit{recovery variable} regulates and completes the activity cycle by
destabilizing the activity states, and thus leading to recurrent switching.
The activity variable exhibits dynamic hysteresis in which the state of
activity is bi-stable if the recovery variable is held constant.  For example
in neuronal dynamics, the cell membrane voltage determines the state of
neuronal activity.  The state becomes active as the voltage depolarizes.  This
depolarization occurs when the potassium-ion conductance, i.e.~the recovery
variable, falls below a threshold.  In return, the depolarized, active membrane
voltage opens voltage-dependent \K-gates causing the \K-conductance to
increase.  Such increases beyond another threshold cause the active state of
the membrane voltage to destabilize, and thereby to complete the cycle.  While
more detailed models can exhibit other complex types of neuronal activity, such
as sub-threshold oscillations and bursting \cite{Braun1980,
Shilnikov2005,Izhikevich2007, Wojcik2011, Shilnikov-12}, this mechanism of
hysteresis, which guides the relaxation oscillations, is retained.

Coupling among relaxation oscillators is introduced through interactions of
their activity variables.  Prominent biological examples of coupled systems
include neuronal populations connected by chemical synapses and gap junctions
\cite{Somers1993,Lewis2003, Bennett2004}, as well as single-cell
organisms communicating via signaling molecules \cite{McMillen2002}.  We
distinguish two types of connections: \textit{excitatory} connections promote
and support the active state, and \textit{inhibitory} connections repress the
active state and hold the inactive state of the oscillator.  Note
that neuronal gap junctions and other types of diffusive coupling do not
fall into either of these categories.

Networked oscillators typically show a degree of coordination among their
individual cycles.  Reciprocal excitation between two oscillators leads to
their synchrony, whereas inhibition forces them to oscillate in anti-phase
\cite{Wang1992}.  Moreover, slow inhibition can promote synchrony in two
coupled oscillatory neurons \cite{Wang1992, Lewis2003, Kopell2004, Terman2013},
while fast, non-delayed inhibition allows for synchronous bursting dynamics due
to spike interactions \cite{Jalil2010, pre2012}.  Generation of robust network
rhythms is of particular relevance for neural ensembles that control the
dynamics of motor patterns.  These central pattern generators (CPGs), described
in the next section, largely inspire the research presented in this work.

\subsection{Dynamics of Central Pattern Generators}

Small neuronal networks of interconnected relaxation oscillators have been
identified in a number of invertebrate CPGs \cite{Sakurai2009, Selverston2010,
Katz2011}.  These networks are structurally equal in individuals of the same
species, and show characteristic differences across related species
\cite{Katz2011}.  A function of the network connectivity is to maintain a
single rhythmic activity pattern of the oscillators, and to ensure resilience
of the pattern against disturbances.  Indeed, computationally extensive
modelling studies of a three-node CPG that controls rhythmic motion in the
stomach of lobsters revealed that a wide range of circuit parameters could
produce the same rhythmic output \cite{Prinz2004, Marder2007, Guenay2010}.
This invariance of pattern generation with respect to parameter changes
highlights the robustness of the network structure to stably produce certain
rhythms.

More complex CPGs support multiple rhythms to efficiently perform multiple
functions \cite{Briggman2006} (see Ref.~\cite{Pisarchik2014} for a review on
multistability).  To study the dynamics of these CPGs, the state space of each
circuit configuration needs to be analyzed for multistability and
polyrhythmicity.  Arguably, the brute-force methods used in aforementioned
lobster studies are numerically impractical in this case, and other methods are
therefore needed.  One potentially viable approach relies on techniques of
qualitative theory to mechanistically understand and categorize the relation
between circuitry and rhythmogenesis \cite{Daun2009, Rubin2009, Dunmyre2010,
Jalil2012, Wojcik2014}.\\

In our recent qualitative studies on circuits consisting of several endogenous
bursters, we have learned that (1)~the duty cycle
\footnote{The duty cycle is the fraction of the active phase divided by the
period, which characterizes bursting neural dynamics.}
of individual neurons strongly affects circuit rhythmicity \cite{Wojcik2011};
(2)~variations in coupling strength among the bursters lead to predictable
bifurcations of rhythms \cite{Wojcik2014}; (3)~strong reciprocal inhibition can
make network rhythms vulnerable to perturbations such as noise
\cite{Schwabedal2014}.  Based on these results, we have been able to better
understand the swim CPG of a sea slug, {\em Melibe leonina}
\cite{Jalil2012,deniz2015}.  Our continuing goal is to develop the theory of
rhythmogenesis to predict changes in rhythm stability, and to determine the
robustness of rhythms in an oscillatory network given its circuitry.

In this paper, we generalize previous results to three-node circuits
constituted of generic relaxation oscillators with relevance outside of
computational neuroscience.  We adopt and counterpose a variety analysis
techniques for polyrythmic circuits and highlight their individual strengths.
Our complimentary techniques of phase reduction, bifurcation theory,
perturbation theory, and stochastic dynamics lets us describe a near-maximal
range of dynamical regimes including different separations of time scale
between activity and recovery variables, bifurcations in individual nodes, and
a range of coupling strengths from weak to strong.  We find that stability and
robustness of circuit rhythms strongly depend on the parameters of oscillators
in the circuit.  In the vicinity of their individual bifurcations, abrupt
changes in the circuit dynamics are observed.  Furthermore, we demonstrate that
qualitative and quantitative stability of polyrhythms do not align completely,
thereby leading us to new hypotheses concerning polyrhythmic circuits.  Our
results strengthen and generalize previous findings obtained with
Hodgkin-Huxley-type neuronal circuits.  Moreover, synthesizing the outcome of
various analysis techniques allows us to specify conditions under which an
individual technique may efficiently extract particular dynamical features of
stability for a given polyrhythmic circuit.\\

Note that resilience of the circuit to retain a rhythm under external
disturbances is analyzed in two ways, in this article, using the terms
stability and robustness.  Stability describes the response of the circuit
dynamics to infinitesimal perturbations.  Robustness describes this response to
finite, stochastic perturbations.

\section{Methods}
\label{sec:methods}

\subsection{Circuit Dynamics}

\begin{figure}[t] \centering
	\includegraphics[width=0.99\linewidth]{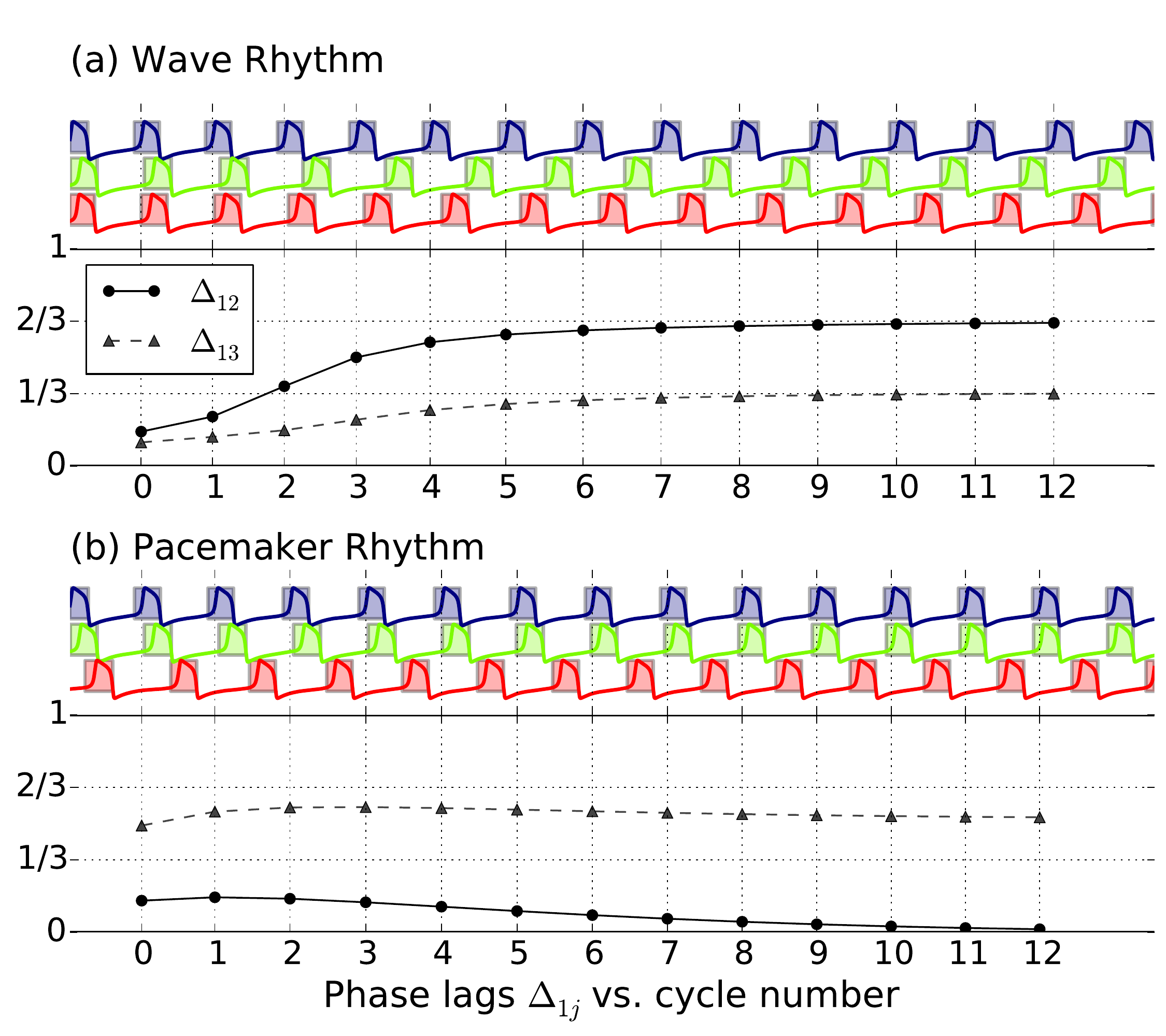}

\caption{(Color online) \textbf{Rhythms in the three-node circuit.}
Oscillations of $V$-variables (top traces) converge to one of the wave rhythms
(a), and pacemaker rhythms (b), both simultaneously stable.  The convergence is
visible in the dynamics of phase lags, $\Delta=(\Delta_{12},\Delta_{13})$, that
approach the characteristic fixed points $\Delta=(2/3,1/3)$, and $(0,1/2)$ in
panel~(a) and~(b), respectively.  Boxes denote active states.  Parameters:
$g=0.005$, $I=0.4$, $\eps=0.17$.}

\label{fig:phasedifference} \end{figure}

We consider a circuit of three Fitzhugh-Nagumo like relaxation oscillators that
are identical and coupled all-to-all.  The circuit dynamics are governed by the
following equations:
\begin{eqnarray} \begin{aligned} \dot{V}_i &= V_i-V_i^3+I-x_i-g\sum_{i\neq
		j}G(V_i, V_j)~,\\ \dot{x}_i &=
		\eps\left[x_{\infty}(V_i)-x_i\right]~, \qquad i,j=1,\,2,\,3~.
	\end{aligned} \label{eq:model} \end{eqnarray}
The state of each individual node is described by the activity variable $V_i$
and recovery variable $x_i$.  Node~$j$ is called \textit{active} if $V_j$
exceeds the activation threshold at $V_j=0$.  Active oscillators repress their
coupling partners towards the inactive state ($V_i<0$).  This interaction
stabilizes certain activity patterns in the circuit dynamics as demonstrated
with two sample trajectories in Fig.~\ref{fig:phasedifference}.  Starting from
arbitrary initial conditions, the phases of activity in each oscillator realign
and converge to a stable circuit rhythm.

The family of circuits is altered by the parameters, \textit{nullcline shift}
$I$ and \textit{inhibitory strength} $g$ discussed in Sec.~\ref{sec:rec}, as
well as \textit{time-scale parameter} $\eps$.  Small values of $\eps$,
e.g.~$\eps=0.1$, indicate well-separate time scales between the dynamics of $V$
and $x$.

Nodes are coupled via a sigmoidal coupling function 
$$ G(V_i, V_j)=\frac{V_i-E}{1+e^{-100\,V_j}}~,$$ 
with $E=-1.5$.  This choice of $E$ makes coupling inhibitory.  The equilibrium
state of the recovery variable is also represented by the sigmoidal function:  
$$x_{\infty}(V)=\frac{1}{1+e^{-10\,V}}~.$$

In a neuroscience context, Eq.~(\ref{eq:model}) represents a phenomenological
model of neuronal dynamics \cite{Fitzhugh1961, Franci2013}.  Variables $V_i$
and $x_i$ describe the membrane voltage and a voltage-dependent \K-conductance,
while the coupling function models fast inhibitory ($V_j>E$) synapses
\cite{Somers1993}.

\subsection{Release, Escape, and Coupling}
\label{sec:rec}

\begin{figure}[t] \centering
	\includegraphics[width=0.99\linewidth]{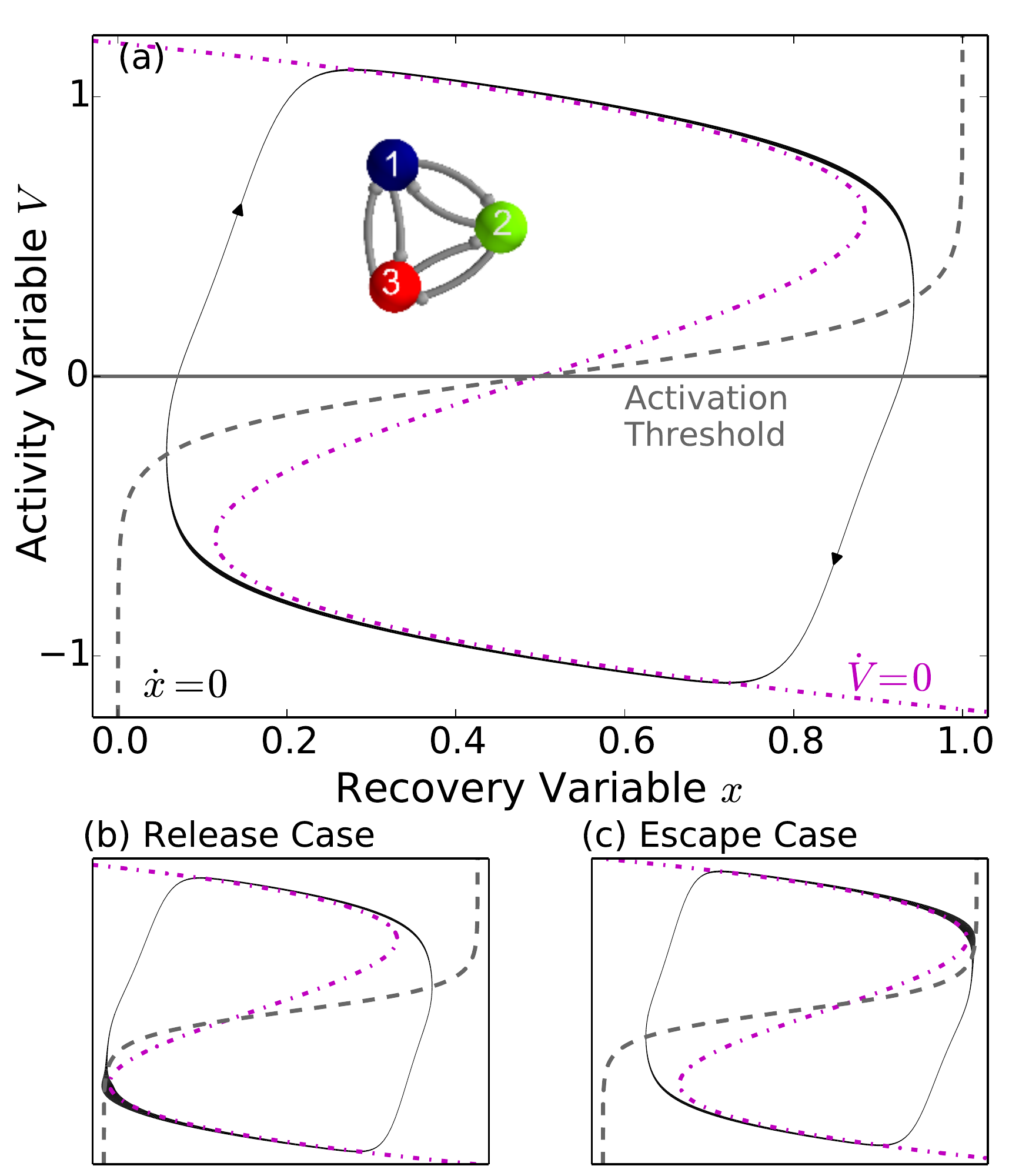}

\caption{ (Color online) \textbf{Dynamics of individual nodes.}  Individual
dynamics $(V_i,x_i)=(V,x)$ (cf.~Eq.~(\ref{eq:model})) of each node in the
circuit (sketch inset in (a)) is governed by the geometric relation of the nullclines
$\dot{x}=0$ (dashed line), and $\dot{V}=0$ (dashed-dotted).  (a) In a regular
configuration at $I=0.5$, the limit cycle (black line) has a very homogeneous
velocity, indicated by its thickness.  When the nullclines almost touch at the
lower (b, $I=0.4$), or upper fold (c, $I=0.59$), the limit cycle stagnates in a
vicinity of the almost-tangency.  These configurations, called \textit{release}
and \textit{escape} cases, can affect the whole network dynamics.  Two arrows
in (a) indicate the orientation of the limit cycle.  The activation threshold
(grey horizontal line) divides active (above) from inactive (below) states.
The inset in panel~(a) shows a sketch of the mutually inhibitory three-node
circuit. Parameters: $\epsilon=0.1$.  }

\label{fig:nullclines} \end{figure}

We investigate the circuit dynamics with individual node dynamics set at a
range of parameters in between two bifurcations controlled by $I$.  Let us
describe individual dynamics at these bifurcations in the $(V,x)$-plane with a
special emphasis placed on the two \textit{nullclines}
\footnote{
The nullcline of a variable $y$ is the set of points defined through
$\dot{y}=0$.}:
the $V$-nullcline is the cubic parabola on which $\dot{V}=0$, and the
$x$-nullcline is a sigmoidal graph on which $\dot{x}=0$; both nullclines are
shown in Fig.~\ref{fig:nullclines}.  Nullclines coordinate the dynamics;
whenever the state is above the $x$-nullcline, $x(t)$ increases and
whenever the state lies to the right of the $V$-nullcline, $V(t)$ decreases,
and \textit{vice versa}.
As shown in the figure, this leads to self-sustained relaxation oscillations in
which the trajectory periodically bypasses the lower and upper fold or knee of
the $V$-nullcline.  The equilibrium state located on the middle segment of the
$V$-nullcline is unstable, and is encircled by the stable periodic orbit in the
$(V,x)$-plane.\\

\noindent\textbf{Release.}
Variations of the parameter $I$ shift the $V$-nullcline relative to the
position of the $x$-nullcline, potentially causing bifurcations in the dynamics
of individual nodes.  Decreasing $I$ shifts the $V$-nullcline to the left.
A tangency of both nullclines occurring near the lower
fold of the $V$-nullcline corresponds to a saddle-node bifurcation
(Fig.~\ref{fig:nullclines}(b)).  If the shifted nullclines locally cross twice,
the node has two additional equilibrium states located in the inactive state:
one unstable and one stable.  This causes the oscillations to cease in the
node which becomes permanently inactive.  Inhibition affects the oscillator,
similarly, in shifting the $V$-nullcline to the left.  When inhibitory
strength, $g$, exceeds a critical value, $g\crit$, the same stable equilibrium
states appear and the inhibited oscillator becomes inactive.  We say that while
inhibited, the oscillator is {\em locked down} at a stable inactive state, and
becomes oscillatory again after it is released from inhibition.  This mechanism
of oscillations emerging from a stable inactive state is called
\textit{release}.  The release mechanism based on the saddle-node bifurcation
works for weak coupling if the gap between the $x$-nullcline and $V$-nullcline
at the lower fold is small.  It is thus a combination of $g$ and $I$ that
determine whether the release mechanism is in place in the circuit dynamics.\\

\noindent\textbf{Escape.} Increasing $I$ shifts the $V$-nullcline to the right
and thereby leads to a bifurcation scenario similar to release at the upper
fold of the nullcline (Fig.~\ref{fig:nullclines}(c)).  Past the corresponding
saddle-node bifurcation the active state of the node becomes a stable
equilibrium.  Inhibition, which shifts the $V$-nullcline back to the left, lets
the oscillator escape from this equilibrium towards the inactive state.  This
mechanism of oscillatory dynamics emerging from a stable active state is called
\textit{escape}.\\

At values of $I$ close to these saddle-node bifurcations -- one near
$I\approx0.4$ and one near $I\approx0.6$, respectively -- the trajectory
bypasses the designated folds, slowly \cite{Shilnikov2001}.  Passage times
of these \textit{stagnation regions} can take a substantial part of the
oscillation period.  Each time is inversely proportional to the square-root of
the gap separating the nullclines \cite{Shilnikov2001}.  This square-root law
yields important intuition about the effect of coupling.   Even weak inhibition
can have a large effect on the oscillation period of the inhibited node,
depending on the gap size.  In this case, it is conceptually difficult to speak
about weak coupling.

\subsection{Construction of \poinc Return Maps}
\label{sec:phaseTrajectories}

Due to the oscillatory nature of the six-dimensional Eqs.~(\ref{eq:model}), the
dynamics can be explained with two variables,
$\Delta=(\Delta_{12},\Delta_{13})$, that describe a maximal number of linearly
independent phase differences, synonymously phase lags, between the three
oscillators.  Herein, $\Delta_{ij}$ describes the phase difference between
node~i and j.  The phase difference between node~2 and 3, $\Delta_{23}$, can be
computed from the other two differences.

Following \citet{Wojcik2011}, we compute these phase lags by first detecting
events $t_1^{(k)}$ at which our chosen reference node~1 becomes active for the
$k$-th time, i.e.~$V_1$ increases through $V_1(t_1^{(k)})=0$.  We also detect
the crossings of nodes~2 and~3, $t_j^{(k)}$ ($j=2,3$) that directly follow each
$t_1^{(k)}$.  Next, we compute the time lags between $t_{2,3}^{(k)}$ and
$t_1^{(k)}$.  Normalizing these time lags by the $k$-th period of the reference
node, $T_1^{(k)}=t_1^{(k+1)}-t_1^{(k)}$, yields a trajectory of phase lags
$\Delta^{(k)}=(\Delta^{(k)}_{12},\Delta^{(k)}_{13})$:
\begin{equation} \Delta_{12}^{(k)} = \frac{t_{2}^{(k)}-t_{1}^{(k)}}{T_1^{(k)}}
	\quad \mbox{and} \quad  \Delta_{13}^{(k)}  =
	\frac{t_{3}^{(k)}-t_{1}^{(k)}}{T_1^{(k)}}~. \label{eq:phaseLags}
\end{equation}
Truncated values $\Delta_{1j}^{(k)}$ modulus-one tabulate the return map on a two-dimensional
torus,
\begin{equation} \Pi:~ \left ( \Delta_{12}^{(k)} ,\,  \Delta_{13}^{(k)} \right)
	\to \left ( \Delta_{12}^{(k+1)} ,\, \Delta_{13}^{(k+1)} \right )~,
	\label{map} \end{equation}
which is computed from long phase-lag trajectories starting from a large number
of initial phase lags between the nodes.  As implied above,
$\Delta_{23}=\Delta_{13}-\Delta_{12}$.

Fig.~\ref{fig:phasedifference}(a) illustrates the relation between the
$V$-traces of the oscillators and their phase lags.  As the number of
oscillations progresses, the phase lags converge exponentially to a locked
state with $\Delta^\ast=(2/3,1/3)$, corresponding to a wave rhythm of
consecutive activity with the order $1$-$3$-$2$.  This locked state is a stable
fixed point (FP) of the return map [Eq.~(\ref{map})].  Using the return map
[Eq.~(\ref{map})], one can show that there co-exist several of such rhythms for
the given circuit configuration.  Fig.~\ref{fig:phasedifference}(b) displays
another example trajectory converging to a pacemaker rhythm characterized by a
FP $\Delta^\ast=(0,1/2)$.\\

To explore polyrhythmic circuit dynamics in its entirety and to identify all
stable rhythms, a regular grid of initial conditions
$\{V(\ph_j^{lk}),x(\ph_j^{lk})| j=1,2,3 \text{ and } l,k=1,\dots,n\}$ is constructed, so that
the corresponding distribution of initial phase lags densely covers the torus.
The shown results were computed using grids of size $40$-by-$40$; however, we
also checked our results on grid sizes of $100$-by-$100$.  For each of these
initial conditions, we compute the phase trajectory as exemplified in
Fig.~\ref{fig:phasedifference}.

All initial conditions lie along the stable periodic orbit $(V(\ph),x(\ph))$ of
uncoupled, individual nodes, which is computed for the given node parameters.
Specifically, node~1 is initialized with zero phase: $\ph_1^{lk}=0$.  The other
two are initialized at phase steps $\delta$ along the orbit:
$\ph_2^{lk}=l\delta$ and $\ph_3^{lk}=k\delta$.  We always set the phase of
zero, where the periodic orbit intersects activation threshold from below.

\begin{figure}[h] \centering
	\includegraphics[width=1.05\linewidth]{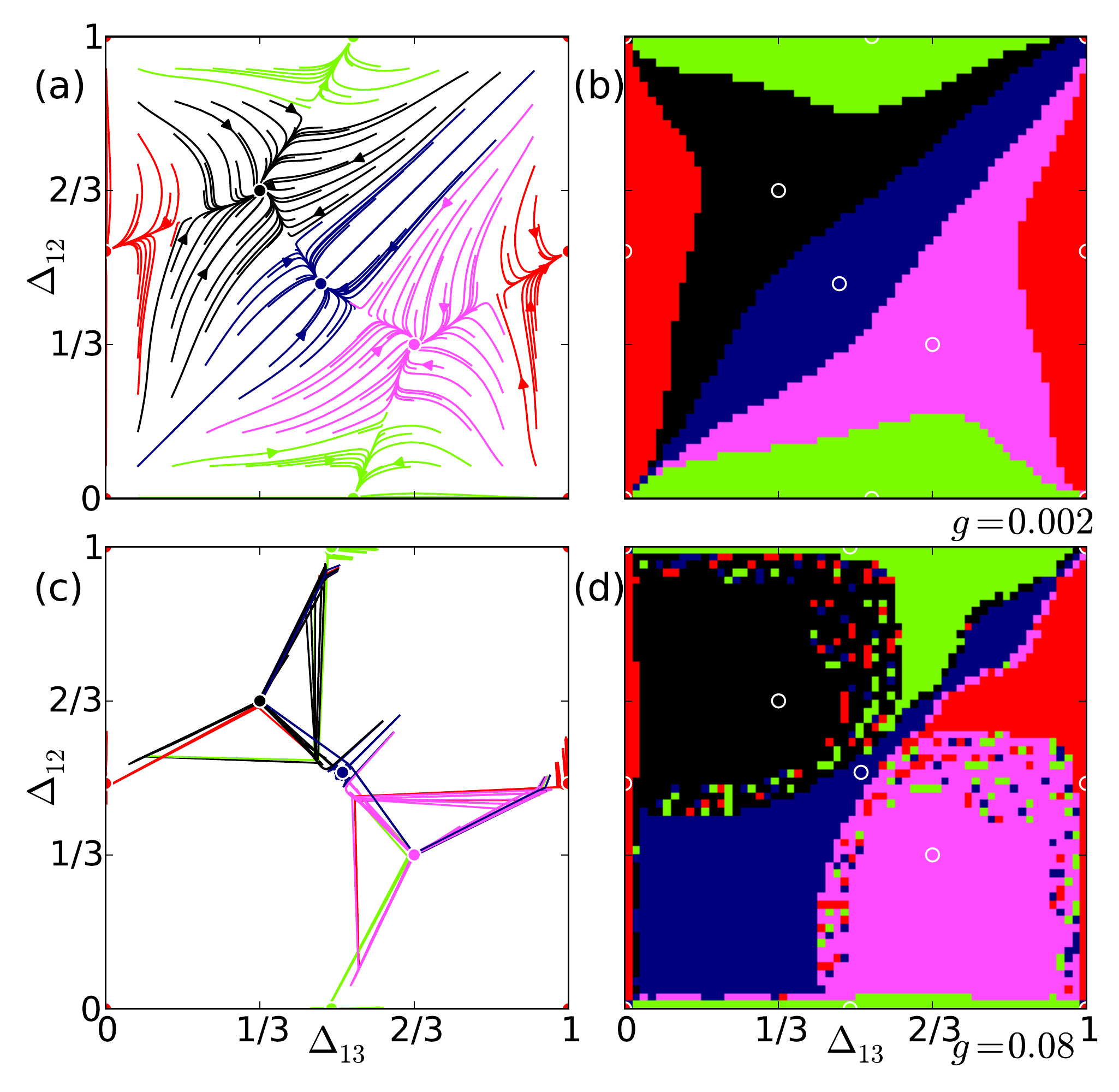}

\caption{(Color online) \textbf{Polyrhythmicity in the torus representation of
phase lags.} \textbf{(a)} At $g=0.002$, the return map reveals the underlying
flow of phases converging to five stable FPs (circles).  \textbf{(b)} By
color-coding initial conditions according to the final rhythm, the basins are
visualized.  \textbf{(c)} At $g=0.08$, convergence in the return map is fast
and no phase flow is discernible.  \textbf{(d)} The structure of color-coded
attraction basins reveals rigid boundaries.  Parameters: $I=0.41$,
$\eps=0.15$.}

\label{fig:torus} \end{figure}

An example for such a phase analysis is summarized in Fig.~\ref{fig:torus}(a),
where we show all phase trajectories on the torus, that were generated from the
grid of initial conditions.
This representation gives the impression of a time-continuous flow of phase
differences, rather than the discrete map [Eq.~(\ref{map})].  All stable and
unstable FPs are visualized as con- or divergence regions: five coexisting
stable FPs are discernible with color-coded attraction basins in the flattened
torus $[0,1)\times [0,1)$.  The coordinates of the stable FPs are associated
	with the locked phase lags of the corresponding rhythms.  We
	differentiate between rhythms of pacemaker and wave type, which we
	identify by their phase lags.  A \textit{pacemaker rhythm} is defined
	to show one phase lag equal zero, and one phase lag close to $1/2$.
	The map in Fig.~\ref{fig:torus}(a) has three corresponding FPs with the
	coordinates in the following ordered pairs: a FP at
	$(\Delta_{12}\approx\frac{1}{2},\Delta_{13}=0)$ shown in red, a FP at
	$(0,\frac{1}{2})$ (green), and a FP at $(\frac{1}{2}, \frac{1}{2})$
	(blue).  Note that, in the latter, $\Delta_{23}=0$.  The other two FPs
	correspond to clockwise and counter-clockwise \textit{wave rhythms}
	defined to show an ordered succession of equidistant phases.  These are
	FPs at $(\frac{2}{3},\frac{1}{3})$ (black) and
	$(\frac{1}{3},\frac{2}{3})$ (pink), respectively.  The attraction
	basins of the FPs are separated by incoming sets (stable separatrices)
	of six saddle FPs not shown in the figure.  Examples of saddles in
	return maps are shown in Figs.~\ref{fig:pitchfork} and
	\ref{fig:waveRhythm_bif}.

\subsection{Mapping Phase Basins}
\label{sec:phaseBasins}

Return maps (Sec.~\ref{sec:phaseTrajectories}) allow us to partition the phase
torus into attraction basins of the coexisting stable FPs.  Numerically, we compute a
basin as a finite set of initial conditions converging to a particular FP.  The
boundaries separating two neighboring basins are approximated by delineating
adjacent initial conditions on the torus that result in two different
rhythms.

To numerically determine these basins, we first create a grid of initial
conditions densely covering the torus.  We then attribute each point on the
grid to a stable rhythm that establishes in the circuit after a transient.
This method is illustrated in Fig.~\ref{fig:torus}(b) showing the torus
partitioned in the five color-coded attraction basins.  For example, all
initial states lying within the blue region converged to the pacemaker rhythm
corresponding to the FP, $(\frac{1}{2},\frac{1}{2})$, which is located in the
middle of the torus.
%The relative size of the attraction basin can be treated as an ``effective
%potential" indicating the likelihood of establishing the given rhythm when
%starting at random.

When coupling is weak, the basin approach does not add information to the
knowledge deducted from return maps.  For example, Figs.~\ref{fig:torus}(a)
and (b) disclose the basins equally well.  This is no longer the case
for stronger coupling, where the approach based on attraction basins becomes very
handy.  For example at $g=0.08$, the phase-basin representation shown in
Fig.~\ref{fig:torus}(d) offers more insight into the circuit dynamics than the
return map shown in Fig.~\ref{fig:torus}(c).  Unlike maps for weak coupling,
stronger coupling results in the faster convergence of $V$-traces and phase
trajectories to corresponding stable FPs over the course of a few iterations.
Therefore, the return-map method gives a good projection of circuit dynamics
only when a slow-fast decomposition is possible.  Herein, the strength of phase
coupling governs the slow time scale, which however, is not small anymore in
the circuit whose dynamics are depicted in Figs.~\ref{fig:torus}(c) and (d).  A
good indication of this limitation is the fractal break-up of basin boundaries
apparent in Fig.~\ref{fig:torus}(d), and which is not observed if time scales
are well separated.

\subsection{Finite Stochastic Perturbations}
\label{sec:stochastic}

\begin{figure}[h] \centering
	\includegraphics[width=0.99\linewidth]{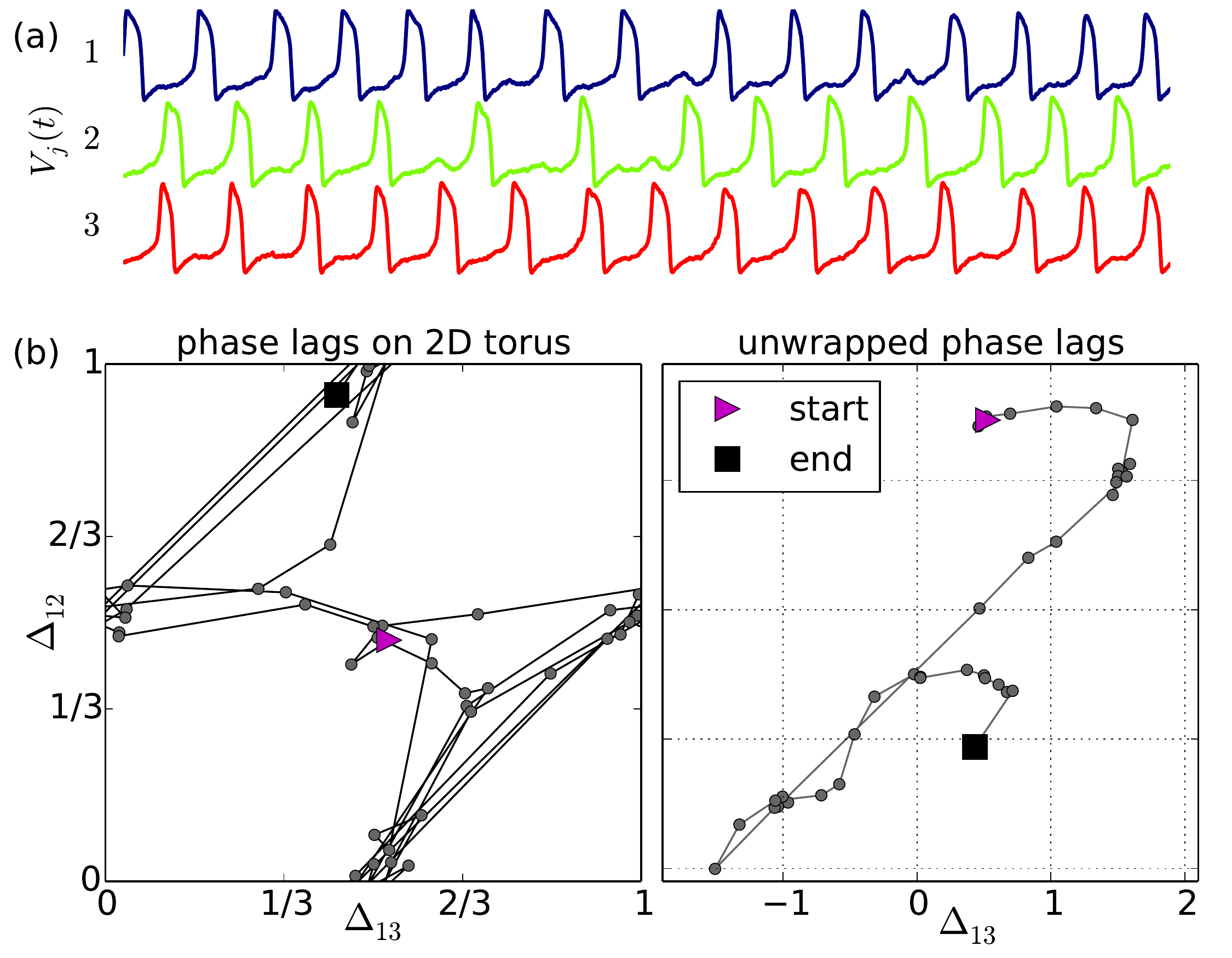}

\caption{(Color online) \textbf{Example of 2D phase diffusion in the stochastic
circuit.}  \textbf{(a)} An excerpt of the example traces $V_j(t)$ shows erratic
transitions between wave and pacemaker rhythms; color-coded bars indicate
synaptic activation in the nodes. \textbf{(b)} Evolution of the corresponding phase lags
(taken on modulo one and unwrapped by Eq.~(\ref{eq:phase})) in a torus (left),
and a phase plane (right) depicting a phase diffusion.}

\label{fig:phD_example} \end{figure}

Robustness of circuit rhythms to perturbations is probed by adding white
noise to each $V$-equation of the circuit in the following way:
\begin{equation} \dot{V}_i=V_i-V_i^3+I-x_i-g\sum_{i\neq j}G(V_i,
	V_j)+\sigma\xi_i(t)~, \label{eq:sde_model} \end{equation}
where $\langle\xi_i(t)\xi_j(t')\rangle =\delta_{ij}\delta(t-t')$ and $i,j=1,2,3$.
Unlike the deterministic case, the stochastic dynamics show sudden
transitions among multiple coexistent rhythms, which are stably generated by
the circuit otherwise.  Such switching is intensified at higher noise intensity
$\sigma$, but foremostly depends on properties of the polyrhythm at given
circuit parameters.  Fig.~\ref{fig:phD_example}(a) illustrates the evolution of
a stochastic trajectory of the circuit that begins in the vicinity of
$\Delta=\left(\frac{1}{2},\frac{1}{2}\right)$.  The circuit mainly switches
among the three coexistent pacemaker states, and sporadically transitions
throughout the wave rhythms.  This wandering of the phase lags is shown in the return
map (Fig.~\ref{fig:phD_example}(b)).  The phase lags, taken modulo one, form
clusters in the vicinity of pacemaker rhythms.  The unwrapped phase lags,
defined without modulo-one and shown in Fig.~\ref{fig:phD_example}(b), allow us
to characterize the two-dimensional random walk.

To evaluate the unwrapped phase lags, we use a more general definition of
phase.  As before, we employ return-time sequences $t_{j}^{(k)}$ for the
oscillators (Sec.~\ref{sec:phaseTrajectories}). Then, define the unwrapped
phase at each return time as follows $\ph_j(t_{j}^{(k)})=k$.  Next, phase
differences, $\ph_j(t)$ are defined in between two successive return times
using linear interpolation ($t_j^{(k)}<t<t_j^{(k+1)}$) \cite{Pikovsky1997}:
\begin{equation} \ph_j(t) =
	\frac{(k+1)(t-t_j^{(k)})+k(t_j^{(k+1)}-t)}{t_j^{(k+1)}-t_j^{(k)}}~.
\label{eq:phase} \end{equation}
This phase differs from the previous definition [Eq.~(\ref{eq:phaseLags})] in
two principal aspects:  (i) the newly defined phase lets us monitor
continuous unwrapped phase differences $\Delta_{1j}=\ph_1-\ph_j$, which cannot
be recovered from the representation with modulo one.  (ii) Through
Eq.~(\ref{eq:phaseLags}) we have phase lags normalized by the period of the
reference node~1, whereas in Eq.~(\ref{eq:phase}), we normalize the
phase of each oscillator with its individual period.  By comparing two panels
in Fig.~\ref{fig:phD_example} one can see that as long as the individual periods
do not differ substantially, both definitions agree well on the torus.\\

\noindent\textbf{Diffusion coefficient.}
The unwrapped phase lags perform a random walk.  After many oscillations, the
variance of each phase lag scales linearly in time:
$\langle(\Delta_{1j}^{(n)})^2\rangle=D_j n$, wherein the proportionality
constant $D_j$ is the diffusion coefficient.  We compute the joint diffusion
constant as a sum, $D=D_2+D_3$, because correlations between phase lags vanish
in the symmetric circuit.  The average is taken over representations of noise.
To estimate $D_j$, we first compute a long phase trajectory $\Delta^{(n)}$.  We
divide the trajectory in segments of $50$ oscillations and compute the two
variances $\langle(\Delta_{1j}^{(n)})^2\rangle$.  Each diffusion coefficient is
determined by a linear fit with respect to $n$, and then the two estimates are
summed to obtain $D$.\\

Alternatively to perturbing $V$-variables, it is possible to add noise
to the $x$-variables.  While the diffusion motion at fixed values of $\sigma$
differs, the qualitative results of polyrhythmic robustness are comparable.

\subsection{Standard Phase Reduction}
\label{sec:PRC}

A perturbation approach lets us derive phase equations for two
phase-difference variables defined on the periodic orbit of the individual
nodes \cite{Kuramoto2003}.   These phase variables can be approximated by those
introduced in Sec.~\ref{sec:phaseTrajectories}.  The computation requires the
uncoupled periodic orbits and their phase resetting curves, which we find with
AUTO (Sec.~\ref{sec:AUTO}).

We map the uncoupled ($g=0$) periodic orbit $y(t)=(V(t),x(t))$ of period $T$ to
a phase variable $\ph(y)\in[0,1)$.  The phase is required to increase
	constantly: $\dot{\ph}=\w=1/T$.  The last assertion fixes the
	definition of $\ph$ up to a constant phase shift.  To quantify how
	coupling influences this phase, we compute the infinitesimal
	\textit{phase resetting curve} $Q(\ph)$ (PRC) for the $V(t)$-variable
	\footnote{A phase resetting curve determines the phase shift an
	oscillator experiences upon an infinitesimal pulse.}.  Then, the phase
	variable $\ph_j$ for node~$j$ is given by ($i,j=1,2,3$)
\begin{equation} \dot{\ph_j} = \w+g Q(\ph_j)\sum_{i\neq j}G(V_j(\ph_j)~,
	V_i(\ph_i))~.  \label{eq:phaseModel} \end{equation}
This already reduces the number of equations from six to three.  Below we will
use a short notation $G(\ph_j,\ph_i)$ for $G(V_j(\ph_j), V_i(\ph_i))$.

For sufficiently small $g$, the phase equations [Eqs.~(\ref{eq:phaseModel})]
hide a separation of time scales allowing for a further reduction to two variables.
The separation becomes visible in Fig.~\ref{fig:phasedifference} showing slow
convergence to stable fixed points over the course of many oscillations.  Therefore, we
consider the phase differences,
$\Delta=(\Delta_{12}=\ph_1-\ph_2;~\Delta_{13}=\ph_1-\ph_3)$.  Their dynamics are 
of the order of the coupling strength $g$, which is slow compared to $\w$:
\begin{eqnarray}
	\begin{aligned}
	\dot{\Delta}_{1j} &= g f_{1j}(\Delta; \ph_1)~, \quad \text{where} \\
	f_{1j} &= Q(\ph_1)\sum_{i\neq 1}G(\ph_1,\ph_i)-Q(\ph_j)\sum_{i\neq j}G(\ph_j,\ph_i).
	\end{aligned}
	\label{eq:phd}
\end{eqnarray}
Introducing $\ph_{i}=\ph_1-\Delta_{1i}$ for $i=2,3$, and
integrating Eq.(\ref{eq:phd}) over the fast variable $\ph_1$ on $[0,1)$, we obtain
\begin{eqnarray}
	\begin{aligned}
		\dot{\Delta}_{1j} &= g f_{1j}(\Delta), \quad \text{where}\\
		f_{1j}(\Delta) &= \int\limits_0^{1}f_{1j}(\Delta;\ph)~d\ph .
	\end{aligned}
	\label{eq:avg_phd}
\end{eqnarray}
This calculation gives us a direct access to the vector field $f_{1j}(\Delta)$
determining all fixed points of the circuit and their stability in the weak
coupling limit.  Moreover, the result depends on neither the choice of $\ph_1$
as the reference phase, nor which phase variable is used for averaging.

\subsection{Bifurcation Analysis and Continuation of Periodic Solutions}
\label{sec:AUTO}

\subsubsection{Stability of Periodic Orbits}

The circuit dynamics [Eq.~(\ref{eq:model})] can also be analyzed with numeric
parameter continuation if circuit rhythms are period orbits \cite{Beyn2002}.
Essentially, one computes the stability multipliers corresponding to a \poinc
map of the orbit.  Let us treat the circuit as a system of six ordinary
differential equations $\dot{y}=f(y;p)$ with a vector $p$ of bifurcation
parameters.  An observable circuit rhythm is a stable $T$-periodic orbit,
$y(t+T)=y(t)$, of this system.  Formal linearizing the system on the orbit
leads to the following variational equation 
\begin{equation} \dot{v}=A(t)v, \quad \text{where} \quad A(t)=Df(y(t);p)~,
	\label{eq:floquet} \end{equation}
with a $6\times 6$ matrix $A(t)$ of periodic coefficients.  This equation
describes how infinitesimal deviations $\xi(0)$ from the periodic orbit may
grow or decay as time progresses: $\xi(t)=\Psi(t)\,\xi(0)$; here  $\Psi(t)$ is
the fundamental matrix \cite{Shilnikov2001}.  Its eigenvalues, $\lambda_k$
($k=0,\dots,5)$, are called the Floquet multipliers.  For each $\lambda_k$,
there is an eigenvector $v_k(t)$ with the property $v_k(T)=\lambda_k v_k(0)$.
In other words, the multiplier $\lambda_k$ quantifies the growth rate of a
perturbation from the periodic orbit in the direction $v_k$ after a single
evolution of the circuit rhythm.

Each orbit always has one multiplier, say $\lambda_0$, equal to $+1$.  It
corresponds to perturbations along the orbit, which neither increase nor
decrease on average over the period $T$.  If all other multipliers fulfill the
condition $|\lambda_k|<1$, then the periodic orbit is Lyapunov stable.  The
values of three multipliers, say $\lambda_3$, $\lambda_4$, and $\lambda_5$, are
close to zero at the coupling strength considered here.  They
correspond to strongly stable directions towards the stable periodic orbits in
the individual nodes.  These directions are perpendicular to those
determined by the vector tangent to the periodic orbit, and hence to those on
which the phase lags are defined on the periodic orbit.  The remaining two
multipliers, $\lambda_1$ and $\lambda_2$, correspond to perturbations parallel
to the phase lags.  These multipliers govern the stability of circuit rhythms,
and are therefore called control multipliers below.

We say that a bifurcation in the circuit occurs when one, or both,
multipliers $\lambda_{1,2}$ cross a unit circle outward, i.e
$|\lambda_k|=1$, as circuit parameters are varied.  This bifurcation gives rise
to the stability loss of a circuit rhythm though a pitchfork or a flip (period
doubling) bifurcation, or the disappearance of the stable rhythm through a
generic saddle-node bifurcation.  The case where a pair of complex conjugate
multipliers $\lambda_{1,2}$ leaves a unit circle corresponds to a
torus or a secondary Andronov-Hopf bifurcation.  This bifurcation can give rise
to the emergence of a stable invariant circle in the \poinc return map.  Such a
stable circle is attributed to the onset of phase jiggling in the voltage
traces \cite{Wojcik2011, Wojcik2014}.
The jiggle frequency is determined by the angle $\tht$ of the
multipliers, i.e.~$\lambda_{1}=e^{\pm i \tht}$. Moreover, if $\tht$ is a simple
multiple of $\pi$, the torus bifurcation unfolding becomes more complex because
of the occurrence of strong resonances at $\tht=\pi$,
$\frac{2\pi}{3}$, and $\frac{\pi}{4}$ \cite{Shilnikov2001}.

\subsubsection{Numerical Computation}

The bifurcation analysis of periodic orbits in the circuit was carried out with
use of parameter continuation package AUTO-07p \cite{Doedel1981}.
Specifically, we set up Eqs.~(\ref{eq:model}) in AUTO to investigate the
stability of the wave and pacemaker rhythms under the variation of parameters
$\eps$, $I$, and $g$.

In our simulations, AUTO was initially used to compute the stable periodic
orbit (PO) and phase resetting curves (PRCs) for each individual oscillator.
Before,  an uncoupled oscillator ($g=0$) was numerically integrated until a
transient relaxed onto the PO, and an individual oscillation was recorded.  The
data was used as an initial guess for AUTO to approximate the PO as precisely as
possible.  By simultaneously solving the adjoint equation, we also obtained the
PRC describing perturbations of the $V$-variable.

Next, AUTO was employed to investigate POs in the full, coupled circuit, to
examine their dependence on the control parameters $I$, $g$, and $\eps$.  We
first found an initial guess for the circuit PO at parameter values $I=0.51$,
$\eps=0.3$, and $g=0.01$.  At these values, the two coexisting wave-rhythm POs
of the circuit are pre-dominantly stable and can be easily detected.  We then
continued either solution in the parameters $I$, $\eps$, and $g$ to examine its
bifurcations, as well as to monitor quantitative variations in the Floquet
multipliers of the circuit PO.  This allowed us to detect bifurcations and
changes in stability.  Similarly, we also investigated properties of the three
symmetric pacemaker rhythms dominating the dynamics of the circuit at $I=0.41$.
Below we analyze in detail the bifurcation boundaries demarcating the stability
and existence regions of the circuit polyrhythm.\\

We performed all of our computations with \textit{Motiftoolbox}, an in-house
developed simulation package that combines powerful computation software
libraries such as Compute Unified Device Architecture, GNU Scientific Library,
python-scipy, python-matplotlib, and AUTO-07p.
\footnote{Motiftoolbox is freely available at
	\texttt{https://github.com/jusjusjus/Motiftoolbox}}.

\section{Results}

\subsection{Qualitative Stability of Polyrhythms}

\subsubsection{Phase Analysis of Polyrhythmicity}

The three-node circuit [Eq.~(\ref{eq:model})] exhibits up to five stable
rhythms, for example at the parameter values $I=0.41$, $g=0.08$, and
$\eps=0.15$, as shown in Fig.~\ref{fig:torus}.  Two rhythms are of the wave
type and three are of the pacemaker type.  While these numbers are formally due
to permutation symmetries of the circuit \cite{Golubitsky2006, Golubitsky2012},
the question of whether the given rhythm exists and is stable or unstable
solely depends on the parameters of the circuit \cite{Wojcik2014}.\\

\noindent\textbf{Return-map analysis.} We investigated the stability of the rhythms for a broad range of circuits by
systematically varying the three parameters $I$, $g$, and $\eps$.  For every
point in a grid of this parameter space, we identified all stable
rhythms by analyzing the phase dynamics of the circuit, as shown in
Fig.~\ref{fig:torus}.  We examined the following parameter ranges of
$\eps\in[0.1,0.3]$, $g\in[0.001,0.1]$, and $I\in[0.4,0.6]$ to span dynamical
scenarios of slow-fast versus normal time scales, weakly versus strongly
coupled circuits, and release versus escape mechanisms of node dynamics,
respectively.
%including soft- and
%hard-locking, release, and escape mechanisms in coupled system
%\cite{belykh2008weak, Rubin2012, Terman2013, Schwabedal2014}. 

\begin{figure}[h] \centering
	\includegraphics[width=0.99\linewidth]{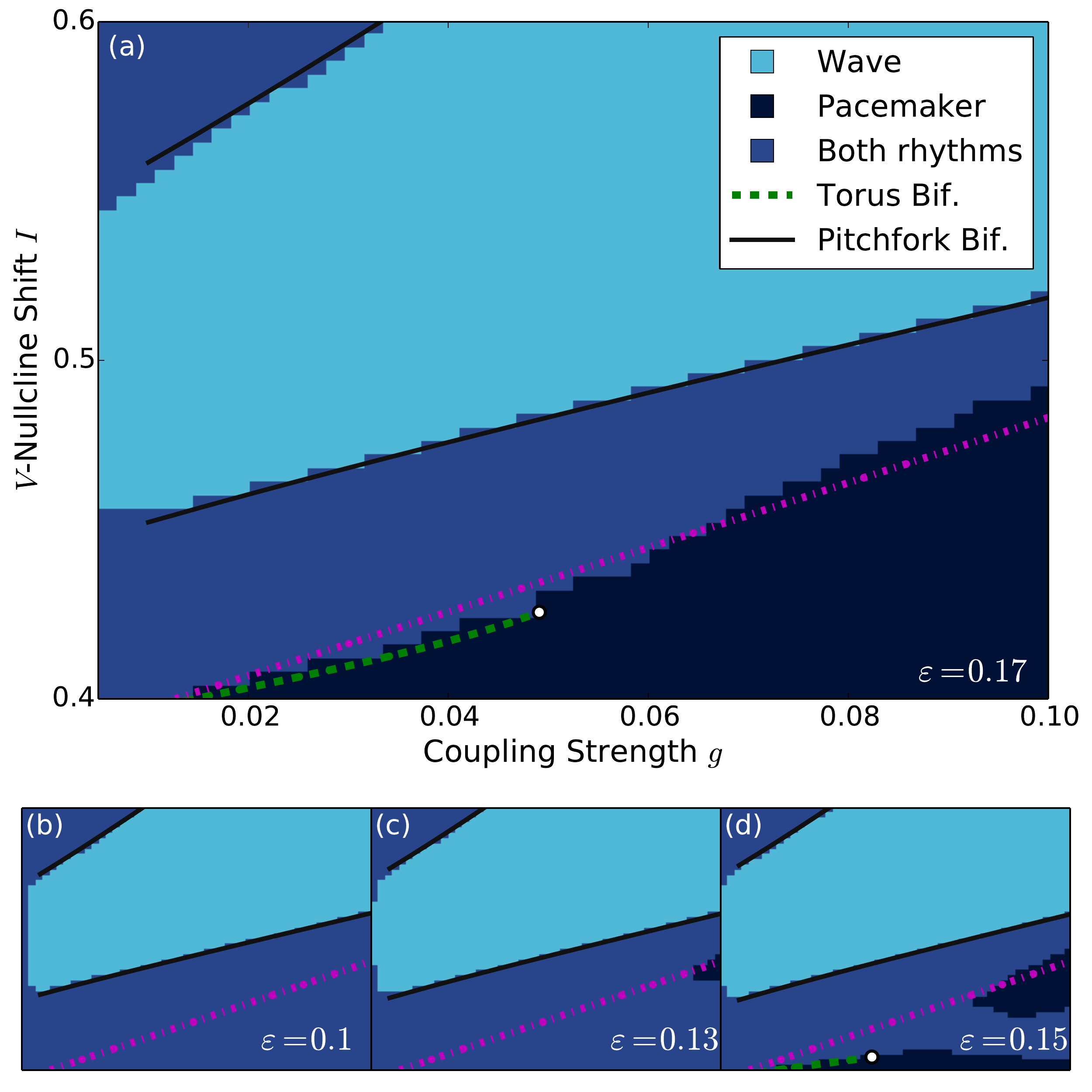}

\caption{(Color online) \textbf{Qualitative stability of circuit polyrhythms in three
parameters.}  For a grid of $I$, $g$ and $\eps$, we determine the stability of
wave and pacemaker rhythms using the phase-basin method
(Sec.~\ref{sec:phaseBasins} and Fig.~\ref{fig:torus}(b),(d)). Each plot~(a)-(d)
shows a bi-parametric sweep in $I$ and $g$ at fixed $\eps$ as indicated.  The
regions of stability for wave (light blue), and pacemaker rhythms (navy) can be
distinct, or show an overlap (dark blue), at which both rhythms are stable.
Pure pacemaker regions at low $I<0.5$ and large $\eps>0.15$ correlate with the
soft-to-hard lock transition (dashed dotted line).  Region borders indicate
bifurcations where one rhythm becomes unstable.  Bifurcation lines of a
pitchfork and a torus bifurcation (Bif.) were determined with AUTO.}

	\label{fig:bifDiag}
\end{figure}

We found regions in the three-dimensional parameter space where either
wave, pacemaker, or both rhythms are stable.  This is illustrated in
Fig.~\ref{fig:bifDiag}, where we show four parameter sweeps in $g$ and $I$,
each one with a different value of $\eps$.

As evident in the figure, we find that the Pacemaker rhythms are stable in a
vicinity of the release and escape case, for which $I\approx0.4$, and
$I\approx0.6$, respectively.  The region of instability, enclosed by solid
black lines, did not seem to depend on $\eps$.  

On the contrary, stability regions of wave rhythms depend on $\eps$.  At values
of $\eps<0.1$, at which time scales are well-separated, the wave rhythms are
stable in the whole parameter space.  At $\eps$ larger than $0.11$, regions in
parameter space form in which wave patterns are unstable.  The dependence is
visible in Fig.~\ref{fig:bifDiag}(c) at $\eps=0.13$, where a region of wave
instability becomes visible at $g>0.08$ and $I\approx0.45$.  The region grows
with $\eps$ and then merges with another region emerging from the release
border at $I=0.4$.

We analytically determined the hard-lock transition of inhibitory coupling
(dashed-dotted [pink] lines in Fig.~\ref{fig:bifDiag}), beyond which one,
active oscillator is able to lock down another oscillator at a stable inactive state.
Because active phases in wave rhythms tend to overlap in time, the mechanism
could influence the stability of these rhythms.  Indeed, we found at large
values of $\eps$, that the boundary of wave stability correlates well with the
hard-lock transition of coupling strength, for example at $\eps=0.17$ shown in
Fig.~\ref{fig:bifDiag}(a).  At smaller values of $\eps$ we did not find such
good correspondence.
%
%\fix\green{(it's not my style to explain a figure in the text.  That's the
%figure caption.)} One can see from Fig.~\ref{fig:bifDiag} that each
%bifurcation diagram is partitioned into three regions: one in light-blue color
%corresponds to the existence region of stable wave-rhythms; black region
%corresponds to the three stable pacemaker rhythms; the above regions are
%separated by a strip (in dark blue) where all five robust rhythms co-exist in
%the dynamics of the circuit.
%
%Pacemaker rhythms are stable in a vicinity of the release and escape case for
%which $I\approx0.4$, and $I\approx0.6$ and $g$ is small, respectively.  At
%intermediate values wave rhythms are stable.
%
%At small values of the time-scale parameter, where the time scales of
%$V$-dynamics and $x$-dynamics are well-separated, the pacemarker patterns are
%generally stable.   At $\eps$ larger than $0.11$, the corresponding regions in
%the bifurcation diagram start \green{to shrink} Fig.~\ref{fig:bifDiag}(a)-(d).
%On the contrary, the stability region of the wave rhythm does not appear to be
%changed \green{by} variations of the parameter $\eps$; \green{it is only}
%determined by the node parameter $I$ and the coupling strength $g$.
%
~\\

\noindent\textbf{Standard phase reduction.}  Return-map analysis is impractical
at weak coupling because convergence of transients to stable rhythms become
very long.  To analyze the stability of polyrhythms in this weak-coupling case,
we used the methods of standard phase reduction (Sec.~\ref{sec:PRC}).

\begin{figure}[h] \centering
	\includegraphics[width=0.99\linewidth]{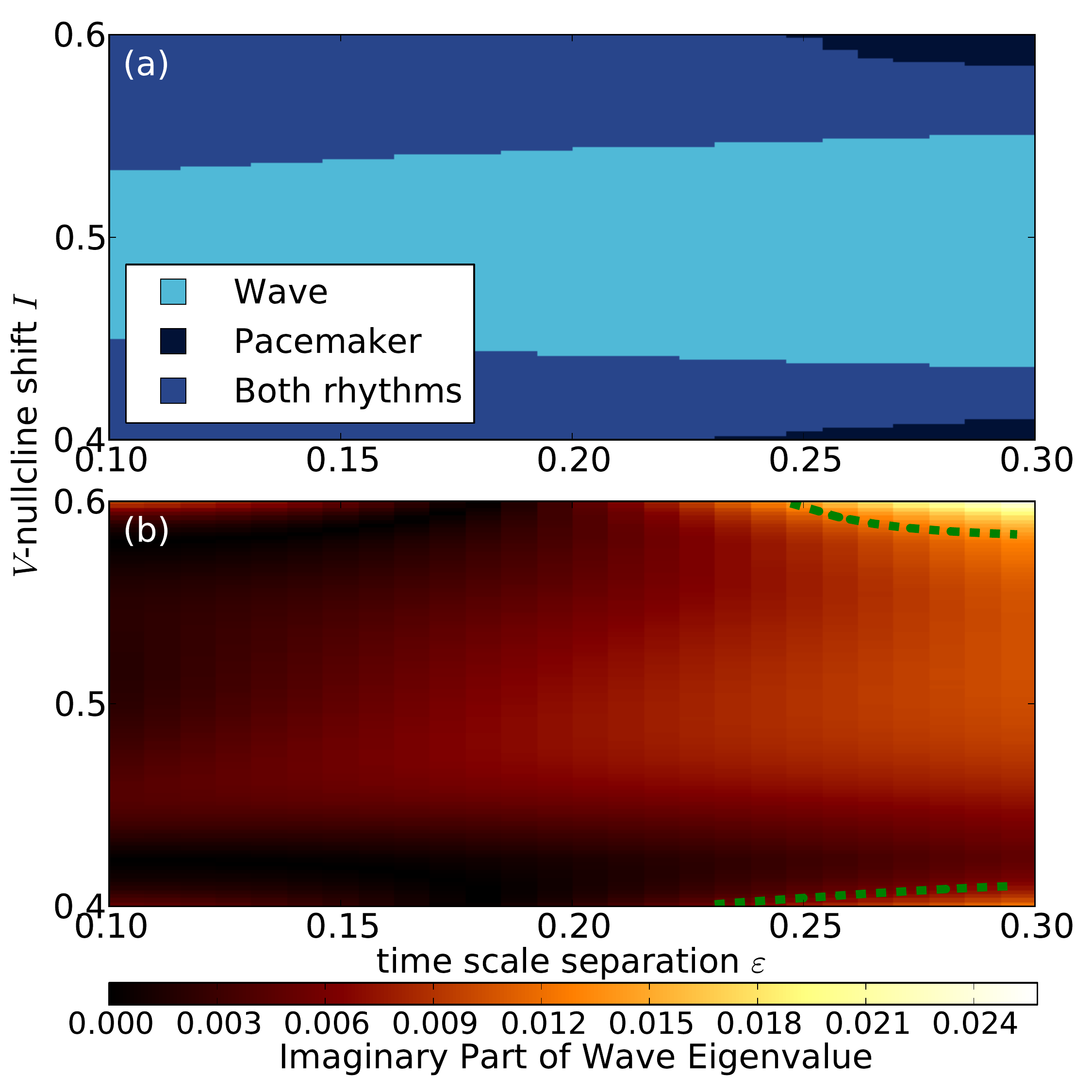}

\caption{(Color online) \textbf{Polyrhythms stability in the weak-coupling
limit.} For a grid of $I$ and $\eps$, we determine the stability of wave and
pacemaker rhythms using the standard phase reduction (Sec.~\ref{sec:PRC}).
\textbf{(a)} a bi-parametric sweep in $I$ and $\eps$ for infinitesimal
$g$.  The regions of stability for wave (light blue), and pacemaker rhythms (navy)
can be distinct, or show an overlap (dark blue), at which both rhythms are
stable.  Region borders indicate bifurcations where one rhythm becomes
unstable.  A torus bifurcation (Bif.) was determined. \textbf{(b)} The
imaginary part of the wave rhythm's Lyapunov exponent is color-coded.  The
exponent is imaginary at the border of wave instability indicating a torus
bifurcation (green dashed line).}

	\label{fig:prcBifDiag}
\end{figure}

We computed phase resetting curves (PRC) for a dense grid of parameters
$I\in[0.4,\, 0.6]$ and $\eps\in[0.1,\, 0.3]$.  From these PRCs, we constructed
the flow for the phase differences on a torus (Eq.~(\ref{eq:avg_phd})).  We
identify all equilibria of this flow.  Applying the numerical differentiation
tools, we can also assess the Lyapunov characteristic exponents of the
equilibria.  Our findings are documented in Fig.~\ref{fig:prcBifDiag}(a)
representing the bifurcation digram in the ($\eps,I$)-parameter plane of the
weakly coupled circuit.

We find that pacemaker rhythms show a region of stability for values of $I$
close to release and escape, while enclosing a region where only wave rhythms
are stable.  This is in line with our results from the return map analysis
(cf.~Fig.~\ref{fig:bifDiag}).  The boundaries of the pacemaker region weakly
depend on $\eps$, such that the region shrinks as $\eps$ increases.  We
note that wave rhythms always coexist for the range of parameter values
in $\eps$ and $I$ considered in Fig.~\ref{fig:bifDiag}. Nevertheless,
at larger values of $\eps\ge0.23$, a region emerges in which the
wave rhythms becomes unstable, and pacemaker rhythms are the
only stable attractors.  Compared to moderate values of
coupling, the regions found here are very thinly around the release and
escape case.
%
%in the circuit due to the release mechanism occurring near the lower fold at
%$I\approx0.4$ (see \ref{fig:nullclines}(b)).  Pacemaker rhythms can also
%emerge in the circuit due to the escape mechanism occurring near the upper
%fold $I\approx0.6$ (see \ref{fig:nullclines}(c)).  These regions can be
%further extended in the diagram for finite coupling strength $g$ (not shown).
%
~\\

The phase-reduced circuit dynamics (summarized in Fig.~\ref{fig:prcBifDiag}) are 
primarily determined by the coupling function $G(V_i, V_j)$ and the phase
resetting curve (PRC) $Q(\varphi)$ given by Eq.~(\ref{eq:phaseModel}).  As $G$
does not depend on the parameters, differences in PRCs are responsible for the
variations in circuit dynamics shown in Fig.~\ref{fig:prcBifDiag}.  We computed
PRCs for a series of parameters $I$ and $\eps$ to understand how the different
patterns of polyrhythmicity relate to these fundamental functions.

\begin{figure}[h] \centering
	\includegraphics[width=0.99\linewidth]{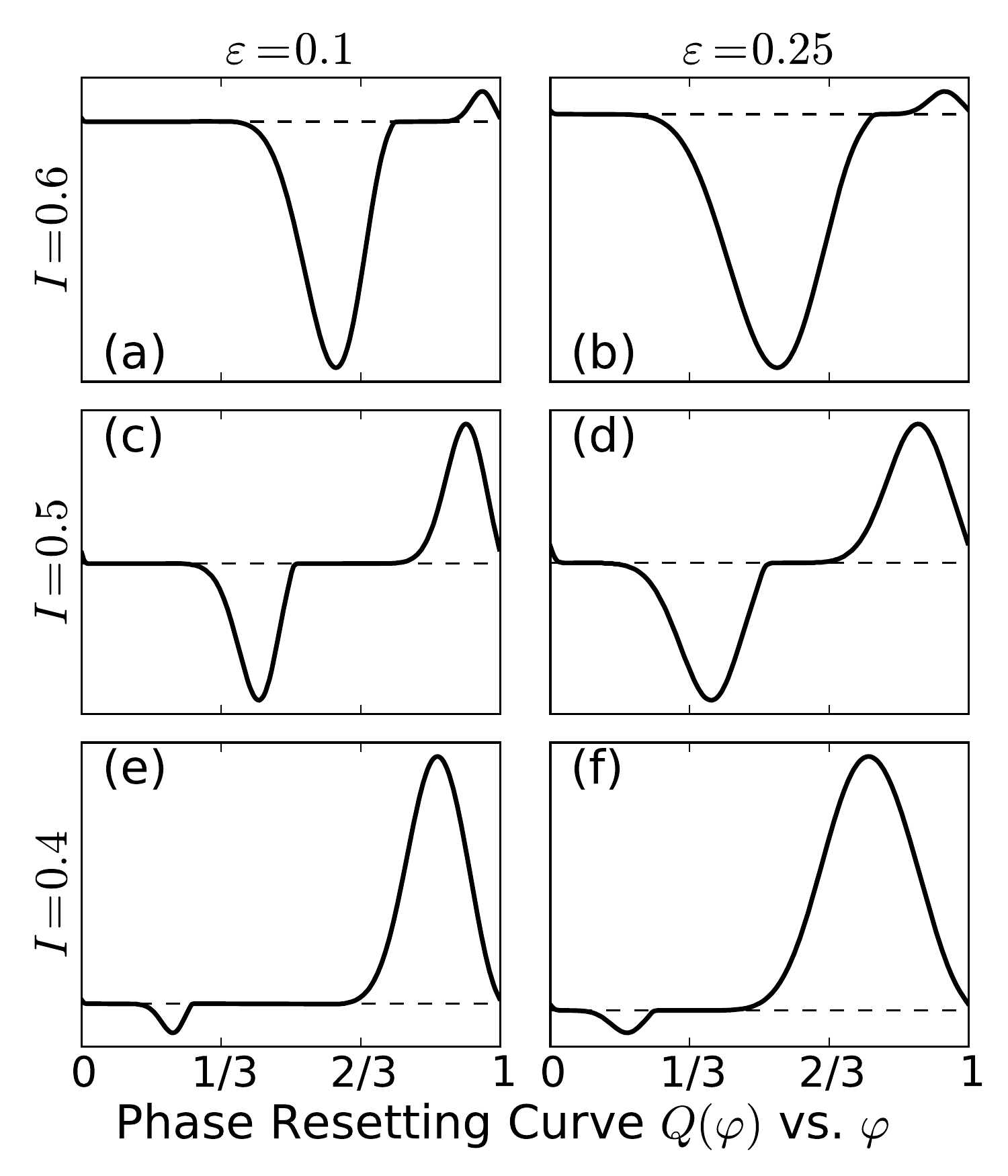}

\caption{(Color online) \textbf{Phase resetting curves of activity variable.}
The infinitesimal phase resetting curve is tabulated for values of $I$ and
$\eps$ (indicated). Values for $I$ correspond to (a, b) escape- ($I=0.6$),
(c, d) normal- ($I=0.5$), and (e, f) release- ($I=0.4$) cases, which are shown for
separate and similar time scales, at $\eps=0.1$ and $\eps=0.25$, respectively.
Note that amplitudes of infinitesimal PRCs are unit-free because perturbations
are linearized.}

	\label{fig:prcExamples}
\end{figure}

Six representative PRCs are shown in Fig.~\ref{fig:prcExamples} for the escape
case in Panels (a, b); for the centered nullclines in Panels (c, d); and for
the release case in Panels (e, f), at two different values of the time-scale
parameter $\eps=0.1$ and $0.25$.  Each curve was characterized by a negative
and a positive inflection, or bump.  These inflections appeared in each of the
phase-parameterized stagnation regions at the lower and upper fold of the
$V$-nullcline.  The upper stagnation region appeared first in the PRC.  The
associated negative inflection indicates that a perturbation with positive sign
causes a phase delay, thus prolonging the active state.  The opposite is true
for the latter, positive inflection that was located at the lower stagnation
region.  Note also that the PRC amplitudes are arbitrary because the
phase theory is taken to a linear order only.  Parameter $I$ affected
PRCs in two ways:  increasing $I$ shifted the first inflection to later phases.
It also rebalances the inflection amplitudes towards the first one.  The
time-scale parameter $\eps$, on the other hand, affects the width of the
inflections.

\subsubsection{Bifurcation Analysis of Circuit Rhythms}

We investigated the dynamical scenarios through which circuit rhythms loose or
gain stability at the region borders shown in Fig.~\ref{fig:bifDiag}.  Such
bifurcations for the wave and pacemaker rhythms, respectively, were visible in
return maps for small enough values of coupling strength, $g$.  It was also
possible to characterize some of the bifurcations with AUTO.\\

\begin{figure}[h] \centering
	\includegraphics[width=0.99\linewidth]{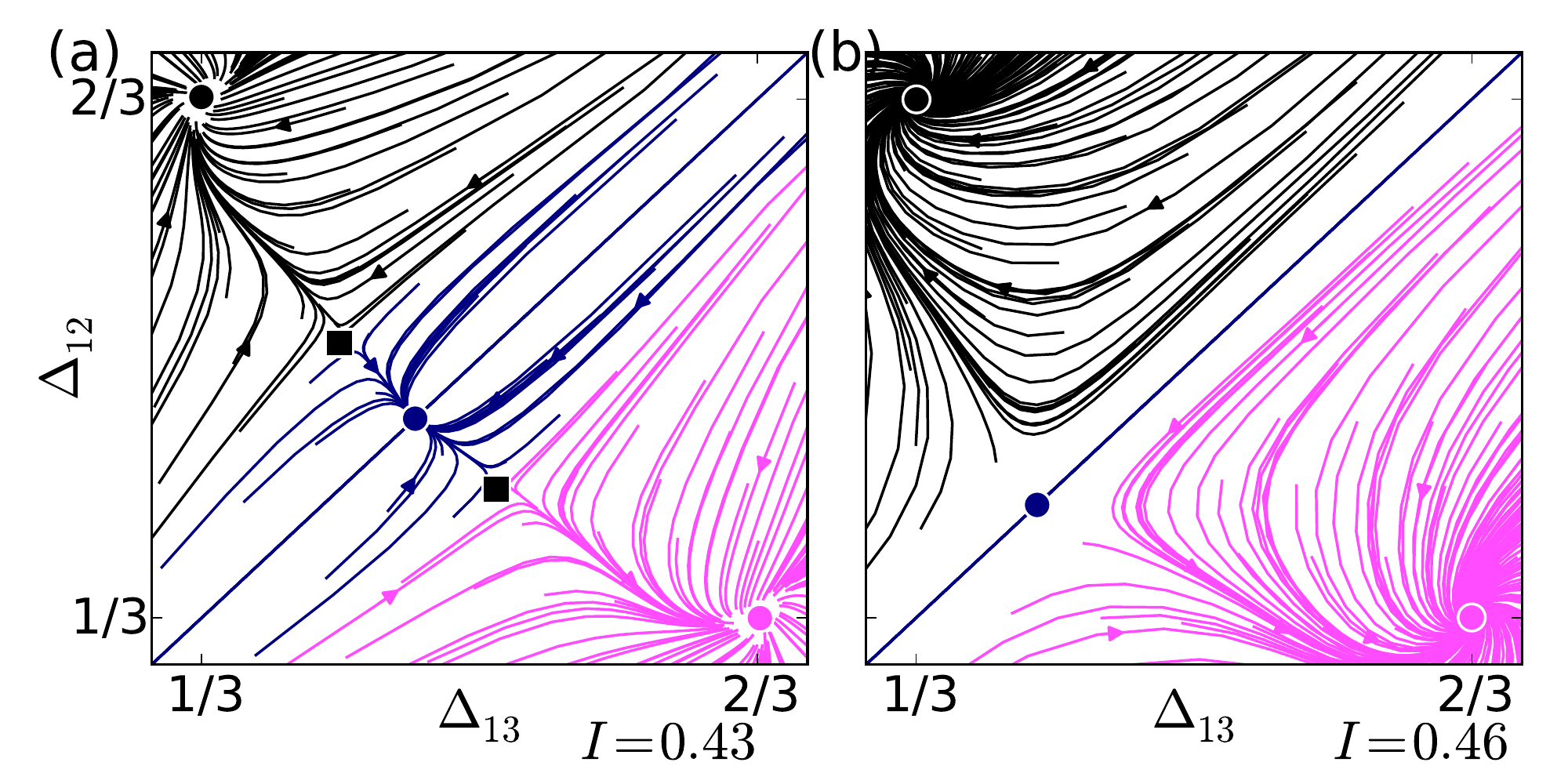}

\caption{(Color online) \textbf{Pitchfork bifurcation of a pacemaker rhythm.}
When $I$ increases from (a) $0.43$ to (b) $0.46$, the pacemaker rhythms undergo
a pitchfork bifurcation, here shown for a part of the full torus
(Fig.~\ref{fig:torus}).  Two saddles (black squares) collide with the fixed
point corresponding to the blue pacemaker rhythm (circle in the center).
Beyond $I=0.46$, only the two wave rhythms (circles in upper-left, and
lower-right corners) are stable.  Parameters: $\eps=0.17$, and $g=0.01$.}
		
	\label{fig:pitchfork}
\end{figure}

\noindent\textbf{Pacemaker rhythms.}
Our analysis of return maps revealed that every pacemaker rhythm first
loses, and then gains stability back through a pitchfork bifurcation as the
parameter $I$ is increased from $0.4$ through $0.6$.  At the pitchfork
bifurcation a stable fixed point (corresponding to either pacemaker) becomes unstable
after it merges with two nearby saddle fixed points to become a saddle itself,
that next becomes stable again through a reverse bifurcation.  Such a bifurcation
sequence was clearly visible in the maps for small $g$, as exemplified in
Fig.~\ref{fig:pitchfork}.

We used AUTO to trace the bifurcation in two parameters with high accuracy.
For this, we initialized one of the pacemaker rhythms as a periodic orbit at
$I=0.41$.  The results do not depend on the choice of rhythm because symmetry
ensures that all pacemaker rhythms have the same stability properties.  We
found the numerically precise parameters of the bifurcations by increasing $I$
at fixed $\eps$ and $g$ when the control Floquet multiplier of the orbit
becomes $+1$.  We then continued the bifurcation in $I$ and $g$ through the
entire parameter range (black lines in Fig.~\ref{fig:bifDiag}).  The procedure
was repeated for each $\eps$.  We found that the bifurcation curves precisely
correspond to the region borders found in return maps.

\noindent\textbf{Wave rhythms.}
We found the bifurcation of wave rhythms to be more complex.  Along the border
of instability (blue and dark-blue region in Fig.~\ref{fig:bifDiag}(a)), the
bifurcation changed its type from torus bifurcation at small $g$, to
saddle-node bifurcation involving three saddles at larger $g$.

\begin{figure}[h] \centering
	\includegraphics[width=1.\linewidth]{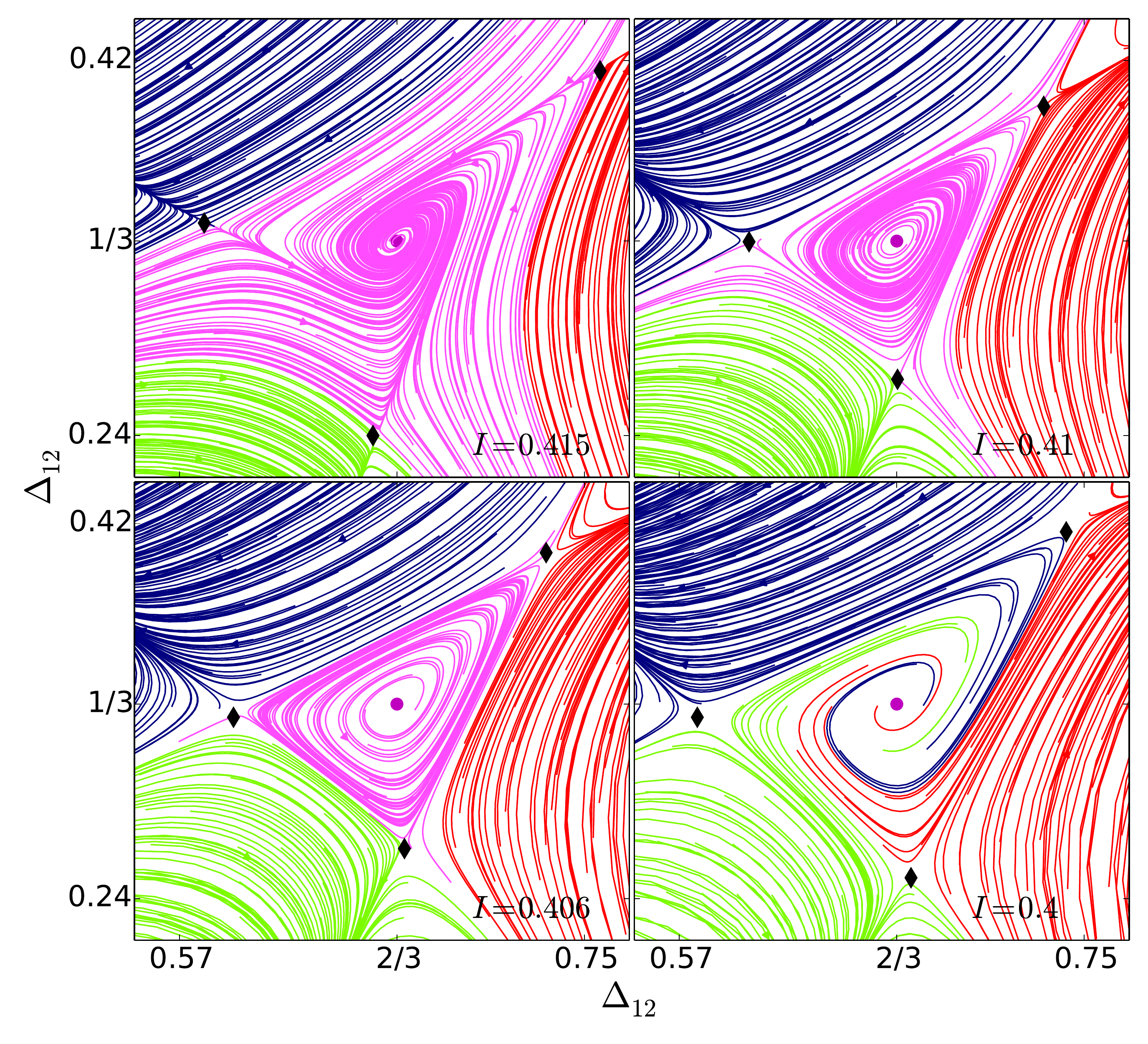}

\caption{(Color online) \textbf{Destabilization of the wave rhythms.} At
	$\eps=0.3$, When we decrease $I$ from $0.415$ to $0.4$ the wave rhythms
	located at $\Delta=(2/3,1/3)$ loses stability.  At values $I>0.415$,
	the basin of attraction of the wave rhythm persistently shrinks.
	Around $I=0.41$, a torus bifurcation gives birth to a stable invariant
	cycle, through which the wave rhythm also loses stability.  The
	invariant cycle grows until it merges with three saddles (black
diamonds) in a heteroclinic bifurcation around $I=0.406$.  After the
heteroclinic bifurcation, the former wave basin is divided among the pacemaker
rhythms. }
		
\label{fig:waveRhythm_bif} \end{figure}

At small $g$, the bifurcation type was discernible in return maps as documented
in Fig.~\ref{fig:waveRhythm_bif}.  Decreasing $I$ at low values of $g$, we
found a torus bifurcation leading to an invariant cycle
(Fig.~\ref{fig:waveRhythm_bif}(a)).  The circle grew in size with further
decreases of $I$ until it became a heteroclinic orbit connecting three saddle
fixed points.  The heteroclinic bifurcation completed the sequence: once the
heteroclinic connection broke down, the pacemaker rhythms dominated the
dynamics of the circuit.
%
%By modifying parameter values, we were able to change the stability of the
%solutions.  Changing $I$ from $0.4$ to $0.415$, at constant values $\eps=?$
%and $g=?$, we found that the two wave solutions changed stability.  The
%recurrence-based phase reduced dynamics for $I$ within that range illucidated
%this change in stability, and disclosed a series of bifurcations which the
%wave rhythm underwent.  At $I=0.415$, the stable rhythm forms focus in the
%phase reduced representation, which loses stability in a supercritical
%Andronov-Hopf bifurcation (cf.~Fig.~\ref{fig:waveRhythm_bif}(d),(c)).  As the
%resultant invariant cycle grows, it simultaneously approches three saddles
%(Fig.~\ref{fig:waveRhythm_bif}(b)).  The resulting heteroclinic bifurcation
%destabilizes the invariant cycle, after which the wave rhythm is unstable.
%
By performing AUTO simulations we could accurately detect the torus bifurcation
in the diagram.  As before, we initiated the circuit on either stable wave
rhythm at $I=0.4$.  The corresponding periodic orbit was then numerically
continued by varying $I$ at fixed $\eps$ and $g$ until AUTO detected the torus
bifurcation.  Next, the torus bifurcation was parametrically continued in $I$ and
$g$, thus tracing down the bifurcation curve represented by dashed lines in the
diagram shown in Fig.~\ref{fig:bifDiag}.  The found segment of the corresponding
bifurcation curve is located in proximity of the associated region border found
through the return maps.  We were not able to detect or continue the
heteroclinic bifurcation.

\begin{figure}[h] \centering
	\includegraphics[width=0.99\linewidth]{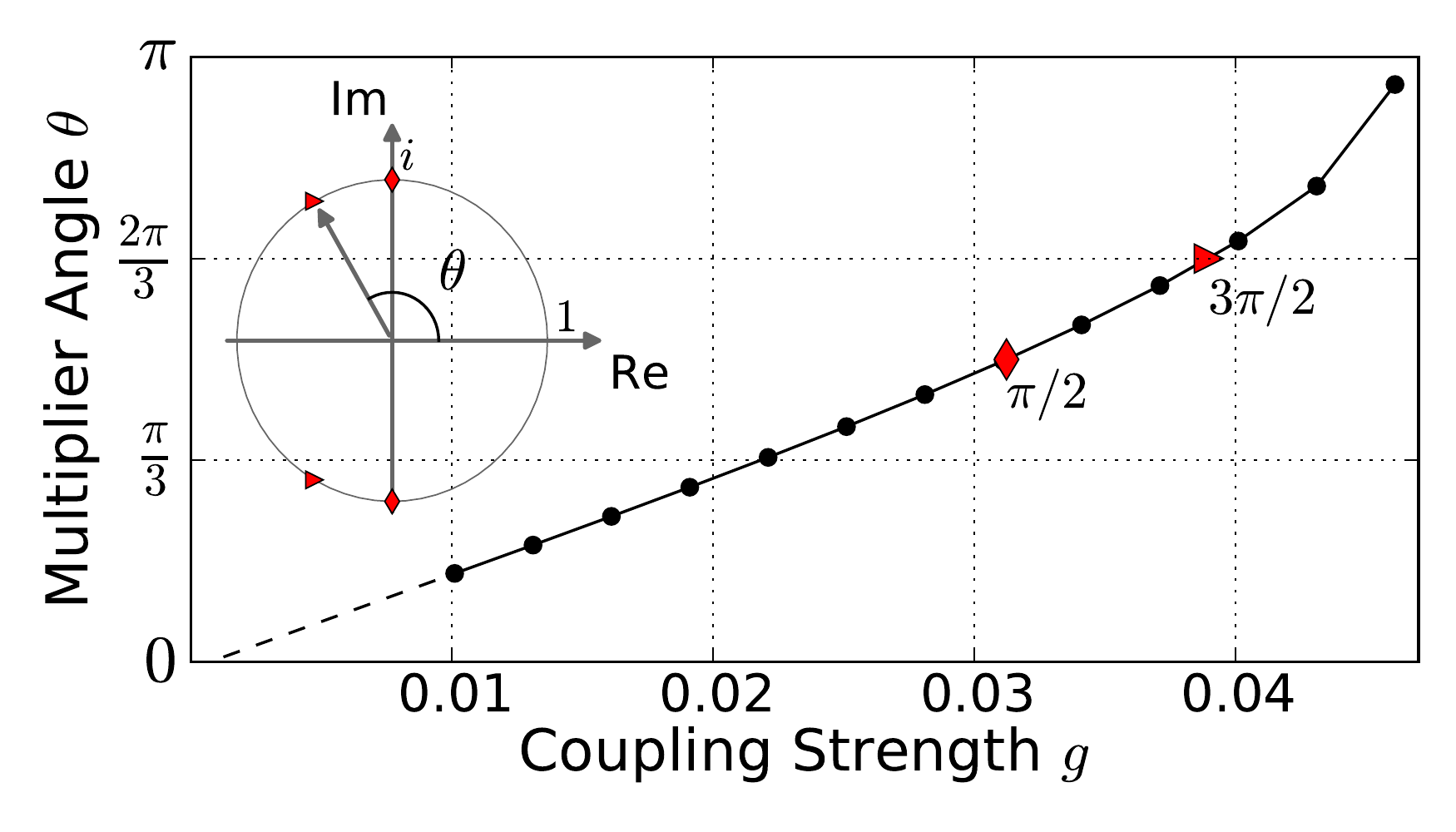}

\caption{(Color online) \textbf{Bifurcation scenario of the wave rhythm.}  The
wave rhythm's complex-conjugate Floquet multipliers $\mu=\e^{i\tht}$ at the
torus bifurcation change their critical angle $\tht$ along the bifurcation
curve (inset).  The angle $\theta$, grows with increasing coupling strength
$g$.  It passes resonances $\pi/2$ and $3\pi/2$ after which $\tht$ approaches
$\pi$.  The angle decreases to zero when extrapolated to $g=0$ (dashed-line).
Parameters: $\eps=0.15$ and $I(g)$ as shown in Fig.~\ref{fig:bifDiag}(d).}
		
	\label{fig:floquetAngles}
\end{figure}

Despite all efforts, we were also unable to continue the torus-bifurcation curve for
values $g$ greater than $0.05$ when using AUTO.  To find out the cause of its
malfunction, we examine the behavior of the pair of complex-conjugate Floquet
multipliers, $e^{\pm i \tht}$, corresponding to the torus bifurcation.  Given
that $||e^{\pm i \tht}||=1$  at the bifurcation, we assessed the angle,
$\tht=|\arg e^{\pm i\tht}|$, as a function of $g$.
In the uncoupled case, the angle is zero because all phase lags are constant.
We found that for increasing $g$ the angle grows monotonically until it reaches
$\pi$, as illustrated in Fig.~\ref{fig:floquetAngles} for $\eps=0.17$.
We also detected the values of $g$ at which the angle reaches strong
resonances.  Of special interest to us is the $\frac{2 \pi}{3}$-strong
resonance. In theory this resonance gives rise to the emergence of a resonant
invariant circle (torus) containing a saddle-node orbit of period three
\cite{Shilnikov2001}.  It is known too that the bifurcation unfolding of the
$\frac{2\pi}{3}$-resonance case involves a further bifurcation resulting in
that   three saddle fixed points collapse into the bifurcating one making it a saddle
with six separatrices.  Studies of such codimension-two bifurcations are the
state of the art that no simulation package can handle.

\subsection{Quantitative Stability of Polyrhythms}

We assess quantitative stability of the circuit rhythms by applying
infinitesimal and finite stochastic perturbations.  Infinitesimal perturbations
do not induce switching from rhythm to rhythm;  instead, they offer insight
into the local stability of wave and pacemaker rhythms separately.  Finite
perturbations induce switching, and therefore inform about the stability of the
polyrhythm, which we also call robustness.

\subsubsection{Linear and Local Stability}
\label{sec:linear}

The standard phase reduction method (Sec.~\ref{sec:PRC}) approximates the phase
dynamics as follows:
$\dot{\Delta} = g f(\Delta),$ ($\Delta=(\Delta_{12},\Delta_{13})$)
The stability of an equilibrium state $\Delta^\ast$ is determined by the
eigenvalues $\lambda_{1,2}$ of the differential $D f(\Delta^\ast)$.  It is
exponentially stable if $|\lambda_{1,2}|<0$.  Coupling strength $g$ scales the
exponent values proportionally: increasing $g$ enhances the stability of a
stable rhythm, while pronouncing the instability of an unstable one.

We counterpose this assertion with the exact calculation of the leading Floquet
exponent $\mu$ using AUTO (Sec.~\ref{sec:AUTO}).  For weak coupling, $\mu$ is
equivalent to the leading eigenvalue $\lambda$ evaluated through phase
reduction.

\begin{figure}[h] \centering
	\includegraphics[width=0.99\linewidth]{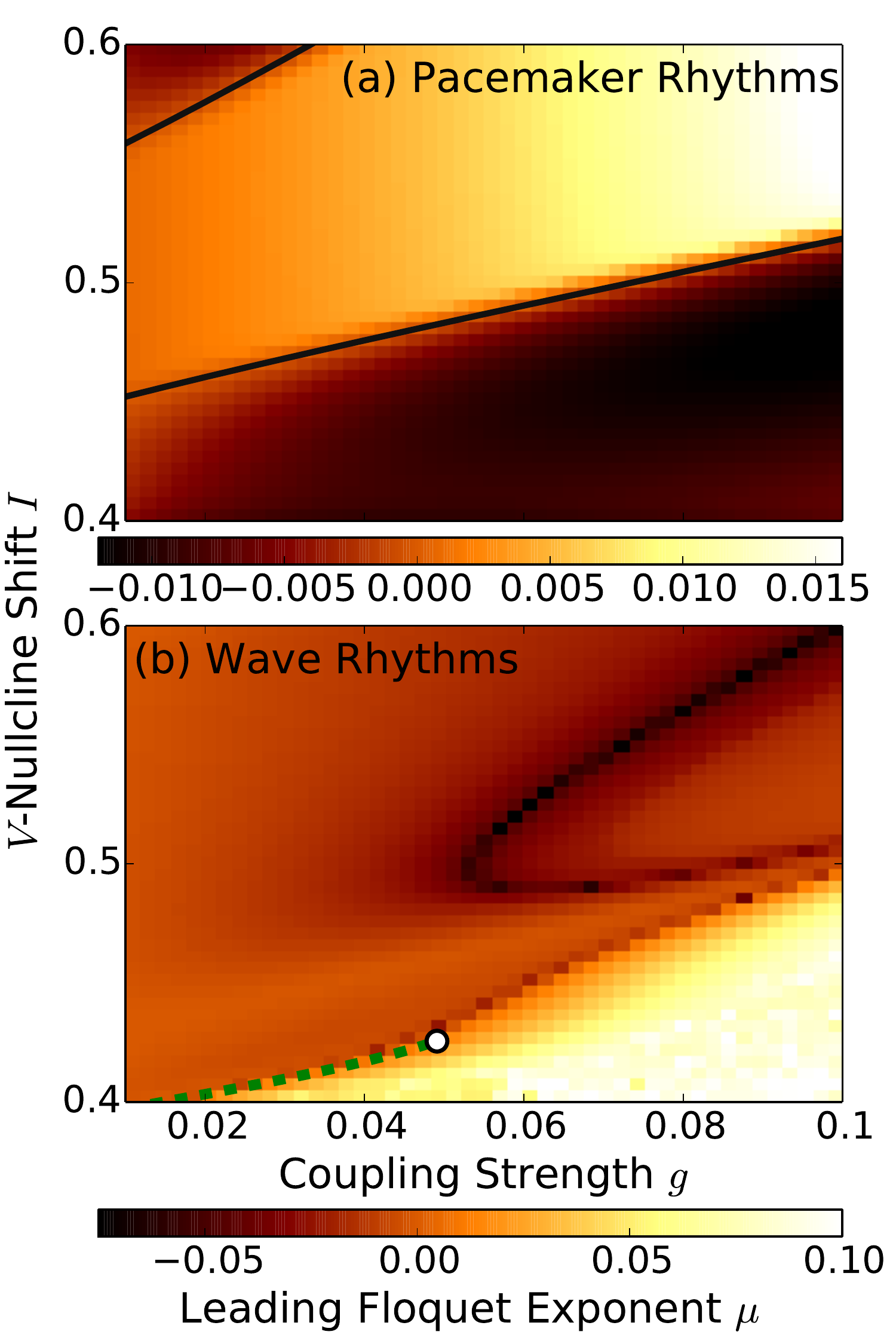}

\caption{(Color online) \textbf{Leading Floquet exponent of circuit rhythms.} The largest,
non-zero Floquet exponent determines the linear stability of the (a) pacemaker
and (b) wave rhythms.  The values, indicating stability ($\mu<0$) or
instability ($\mu>0$) of the rhythms, are proportional to $g$ at weak coupling,
unless in the vicinity of bifurcations (black-solid, and green-dashed lines).
The wave rhythms are the most stable on a parabola-shaped set in the
$(g,I)$-parameter plane, and highly unstable at low values of $I$.}

\label{fig:Floquet} \end{figure}

For a grid of parameter values $I$, $g$ and $\eps$, we computed the leading
Floquet exponent $\mu$  for the traveling wave and a pacemaker rhythm.  Note
that all permutation-symmetric rhythms have the same set of exponents
\cite{Golubitsky2006, Golubitsky2012}.  The corresponding bifurcation diagrams
for $\eps=0.17$ are shown in Fig.~\ref{fig:Floquet}
(cf.~Fig.~\ref{fig:bifDiag}(a)).  The exponents changed signs exactly at the
pitchfork and torus bifurcations for the pacemaker and wave rhythms,
respectively.  For weak coupling strength $g$, the exponent $\mu$ shows a
monotonous dependence on $g$, which breaks down in a vicinity of the
bifurcations.  Strengthening $g$ for the wave rhythm revealed a parabola-shaped
set of minimal values of $\mu$ in the bifurcation diagram.  At these parameter
values, the local linear stability of wave rhythms reached its maximum.  At low
values of $I$, the wave rhythms become highly unstable.

\subsubsection{Robustness of Polyrhythmicity}
\label{sec:stochastic}

We tested the robustness of polyrhythmic circuit dynamics under noisy
perturbations.  Robustness was quantified by the phase diffusion constant for
the phase-lag variables described in Sec.~\ref{sec:stochastic}.  We found that
the diffusion constant varies by orders of magnitude across the tested
parameter space of $I$, $g$, and $\eps$.  Therefore, we carefully selected a
value of noise intensity $\sigma$ that allowed us to sample the wide range of
parameters with comparable accuracy.  At small $\sigma$, noise caused only few
switching events within $20000$ circuit-rhythm periods, and therefore no
feasible estimation for $D$ was possible.  We therefore present our results for
$\sigma=0.01$, below.  The value, $\sigma=0.02$, yielded similar results.

\begin{figure}[h] \centering
	\includegraphics[width=0.99\linewidth]{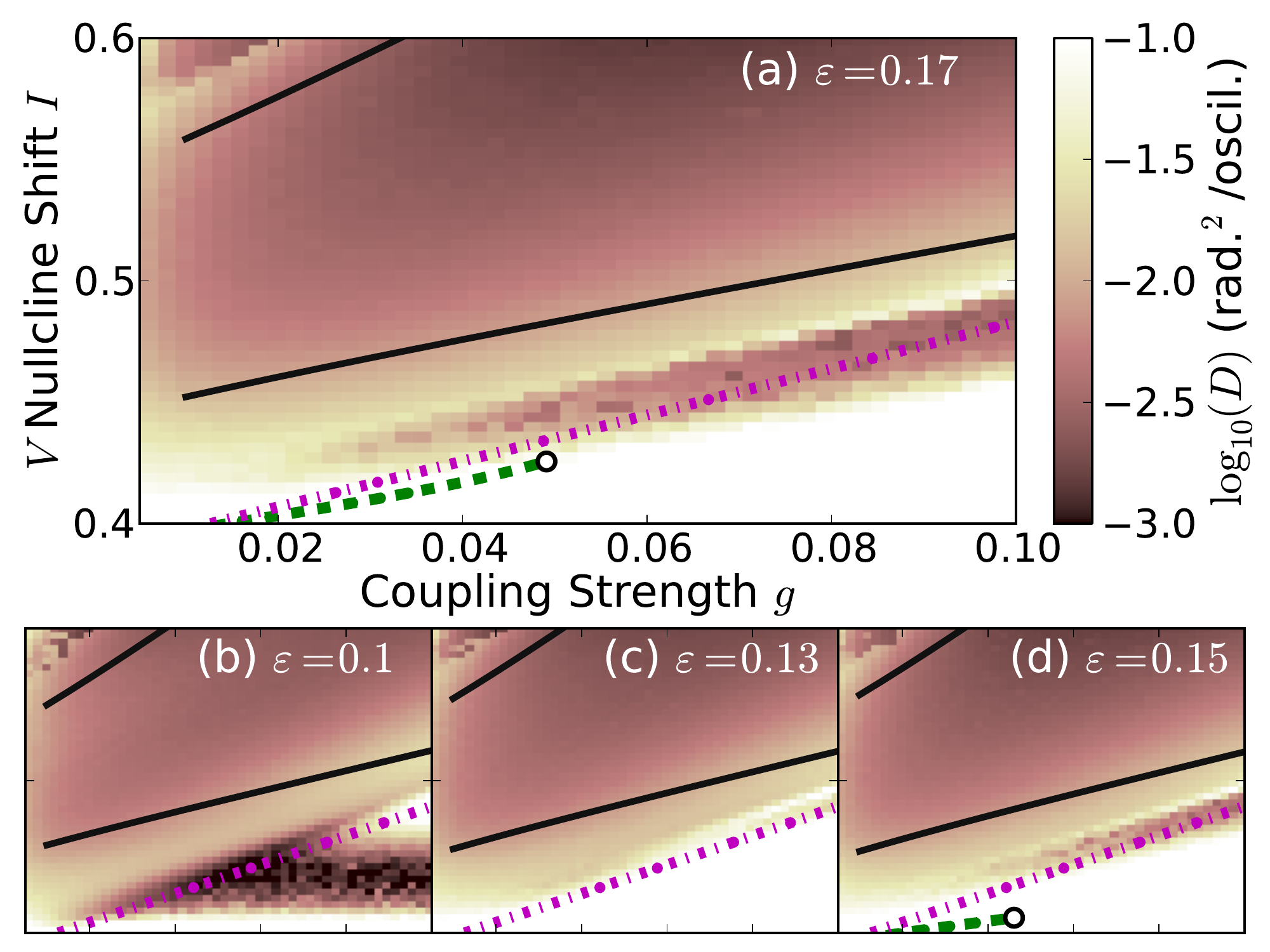}

\caption{(color online) \textbf{Phase diffusion constant $D$ of stochastic
circuit dynamics.}  Random perturbations induce rhythm switching in the
circuit, resulting in a finite phase diffusion constant $D$.  Where the circuit
demonstrates only wave rhythms (between black lines), $D$ depends weakly on
$I$, $g$, and $\eps$.  Elsewhere, the dependence is complex and cannot be explained
by the local bifurcation structures in the coexistence region of all five
polyrhythms.  $\sigma=0.01$.}

\label{fig:D_allfour} \end{figure}

In the region where waves are the only stable rhythms, the phase
diffusion constant showed a monotonous dependence on all parameters
(cf.~Fig.~\ref{fig:D_allfour} between the solid black lines).  Outside this
region,  the parameter dependence of $D$ shows considerable complexity.
In Fig.~\ref{fig:D}, we supplement our findings with exemplary traces of
the circuit dynamics.

Close to the escape case, the phase diffusion constant $D(I)$ at fixed $g$ and
$\eps$ shows a series of minima and maxima.  A comparison with the
deterministic bifurcation diagram (Fig.~\ref{fig:bifDiag}) revealed that the
minima of $D$ somewhat align with both, the soft-to-hard lock transition line
and the wave-instability line at $\eps>0.13$.  However, this is not the case at
$\eps=0.1$ where the wave rhythm did not bifurcate, while $D$ still showed a
pronounced valley of stable dynamics (cf.~Fig.~\ref{fig:D}(1),(2)).

The diffusion constant $D$ becomes increasingly large as $I$ approaches the
boundary $0.4$ for all values of $g$ and $\eps$.  This is related to the
highly vulnerable dynamics of the individual oscillators near the saddle-node
bifurcation.

\section{Discussion}

Previous studies have demonstrated that mutually inhibitory three-node circuits
of neuronal bursters synchronize in up to five coexistent stable rhythms
\cite{Wojcik2011, Wojcik2014, Schwabedal2014}.  Out of these five, three are
pacemaker rhythms, and two are the clockwise and counterclockwise wave rhythms.
Disturbances with external current pulses or noise can cause switching among
the rhythms.  To better understand mechanisms of stability and robustness of
polyrhythmicity, we have explored in this paper circuit dynamics constituted by
generic relaxation oscillations.  We set out to explore a wide parameter range
to catalogue and describe the circuit dynamics in its entirety.

In particular, we investigated the circuit dynamics [Eq.~(\ref{eq:model})]
depending on three principle parameters: the time-scale separation $\eps$
determines the speed of the recovery variable $x$ with respect to the activity
variable $V$ in each node; parameter $I$ shifts the position of the
$V$-nullcline and thus controls the release and escape mechanism
(cf.~Fig.~\ref{fig:nullclines}(b),(c)); and the inhibitory coupling strength
$g$ determines how strong the node dynamics are tied to each other in the
circuit.  Here, we also distinguished a hard-lock coupling regime where
inhibition is strong enough to fix the inhibited node in the inactive state.

We used in our circuit a Fitzhugh-Nagumo model with an $x$-nullcline,
$x_\infty(V)$, that shows a sigmoidal shape and thereby deviates from the
standard linear function.  This choice is relevant especially for biological
applications: it closely resembles corresponding Boltzmann and Hill functions
that appear in Hodgkin-Huxley-type neuronal dynamics and enzyme kinetics, for
example.  Moreover, our choice allowed us to study transitions in the circuit
dynamics related to the release and escape mechanisms, which are fundamental
mechanisms of rhythmogenesis.

\subsection{Qualitative Stability of Polyrhythms}

Circuit dynamics were particularly sensitive to the $V$-nullcline shift $I$
when set close to the release or escape case
(Fig.~\ref{fig:nullclines}(b),(c)).  In both cases, the dynamics of individual
oscillators are close to a saddle-node (SN) bifurcation emerging around the
lower or upper fold of the $V$-nullcline, respectively.  Such SN ghosts are
known to substantially reduce the individual oscillation frequencies.  Our
results show that the two mechanisms qualitatively affect the circuit dynamics
through interactions with inhibitory coupling, as described further below.

Let us first discuss the generic case of intermediate $I$-values.  Here,
individual oscillators are not close to any bifurcations, and we observed wave
rhythms as the only stable circuit dynamics (cf.~Fig.~\ref{fig:bifDiag}).  The
inhibition exerted by an individual oscillator is not strong enough to overcome
mutual phase repulsion of the other two oscillators.  Conversely, the pacemaker
rhythms were unstable in this region.\\

\noindent\textbf{Release case.}
At low values of $I$, corresponding to the release case, inhibitory coupling
has a drastic effect on the dynamics of the inhibited nodes, specifically in
the region of stagnation at the lower fold.  Weak inhibition brings the
dynamics of individual oscillators closer to the SN bifurcation at the lower
fold; inhibition where the coupling parameter is larger than $g\crit$ induces a
transient SN bifurcation, leading to a stable equilibrium state
\cite{Schwabedal2014}.  With such strong inhibition, one oscillator locks down
the other oscillators in the inactive state for the time it is active.  As a
result, the distribution of phases along individual orbits is highly
non-uniform, and condensed around the stagnation region: each oscillator is
either in a short active state, or it stagnates near the SN equilibrium.
Naturally, two of the three oscillators must collapse in one of these two
states, and thus synchronize.  According to this description of two
quasi-discrete states, the three-state wave rhythms are very unstable.
Evidence for this heuristic description is the close proximity of the border of
wave stability (border of blue to navy region in Fig.~\ref{fig:bifDiag}(a)) and
the hard-lock transition line (magenta dashed-dotted line), that was visible at
$\eps=0.17$.  For smaller values of $\eps$ however, i.e.~at large time-scale
separation, we did not observe this mechanism (cf.~Fig.~\ref{fig:bifDiag}(b)
where $\eps=0.1$).  At these values, the slow dynamics of the recovery variable
spreads active and inactive states along the branches of the slow manifold.
Therefore, the heuristic description is not valid in this case.

\noindent\textbf{Escape case.}
When increasing $I$ towards the escape case, we again found a region of
parameter space wherein pacemaker rhythms are stable.  As shown in
Fig.~\ref{fig:bifDiag}, the effect did not depend on $\eps$ and was most
pronounced at small coupling strengths, which we also confirmed in the
weak-coupling limit (Fig.~\ref{fig:prcBifDiag}(a)).  The SN ghost, here located
at the upper fold (Fig.~\ref{fig:nullclines}(c)), plays a key role in the
emergence of pacemaker rhythms.  Analogous to the release case, the upper fold
forms a stagnation region that leads to a highly non-uniform distribution of
phases.  Consider oscillators~1 and 2, that are in active states.  The states
approach each other as they slow down in the vicinity of the stagnation region.
When the third oscillator becomes active, its inhibition breaks stagnation by
widening the gap between $V$- and $x$-nullcline.  Thus, oscillators~1 and~2 can
simultaneously ``escape'' from the ghost and synchronize in a pacemaker rhythm.
With only this mechanism, the pacemaker region should extend to higher coupling
strengths, which is not the case (cf.~Fig.~\ref{fig:bifDiag}).  The mutual
interaction of the oscillators~1 and 2, in the example, prevents their
synchronization for strong coupling: if they strongly inhibit each other in the
stagnation region, the oscillator~1 slightly lagging behind will repress
oscillator~2 and push it through the stagnation region into the inactive state.
In effect, oscillator~2 will cease to inhibit oscillator~1 which, thus, lingers
on in the stagnation region.  This explains our observation that the pacemaker
rhythm was unstable.\\

%\green{The period of individual periodic orbits depends linearly on $\eps$ if
%the oscillator is far from saddle-node bifurcations.}  Increasing $\eps$ also
%shrinks the limit cycle in the state space $(V,x)$ because of the faster
%transitions between the top and low branches of the $V$-nullcline.  At smaller
%$\eps$ values for the given configuration of the slow nullcline $\dot{x}=0$,
%the phase point spends most time near the two folds of the
%individual period orbit where its stability and dynamics becomes vulnerable to
%inhibitory perturbations.  Conversely, at larger $\eps$ values, the recovery
%period, also called the refractory period, of the phase points of the node
%shortens and hence, the  limit cycle of the driven becomes less vulnerable
%timewise to external inhibition.  This suggests that from a viewpoint of Markov
%chains, in this latter case the distribution of three phases would be the
%traveling wave rhythms rather than the pacemaker ones in the formal case at
%smaller values of $\eps$.

The results for the release case are in line with those of \citet{Wojcik2011}.
In their model of neuronal bursting, which closely resembles the neuronal
electrophysiology, a shift of a \K-conductance parameter induces the same
series of bifurcations of the wave rhythms as those shown in
Fig.~\ref{fig:waveRhythm_bif}.  A dynamical analysis revealed that the shift
widens the gap between the slow and fast nullclines at the lower fold
\cite{Schwabedal2014}, thus completing the analogy of the two observations.  On
the contrary, the escape mechanism is not observed due to inhibitory
interactions of spikes \cite{Jalil2010}.  In these parameter regimes, the
burster models show different circuit dynamics compared with our relaxation
oscillator.

\subsection{Quantitative Stability of Polyrhythms}

Functional circuits often operate in environments where perturbations and noise
interrupt their dynamics.  In polyrhythmic circuits, this can lead to switching
between coexistent rhythms and the switching process strongly depends on the
circuit parameters.  To analyze how robustly the circuit sustains a rhythm in
such an environment, we randomly perturbed the circuit dynamics and monitored
how individual phases diffused apart in consecutive random switching events.
We found that the phase diffusion constant, indicating robustness of the
polyrhythm, strongly depended on circuit parameters.  However, we were unable
to predict robustness by the bifurcation structure in the circuit or by
linear-stability measures of individual circuit rhythms.  One may still
speculate why certain circuit configurations are more robust than others based
on these information, for which we give two examples below.\\

\begin{figure}[h] \centering
	\includegraphics[width=0.99\linewidth]{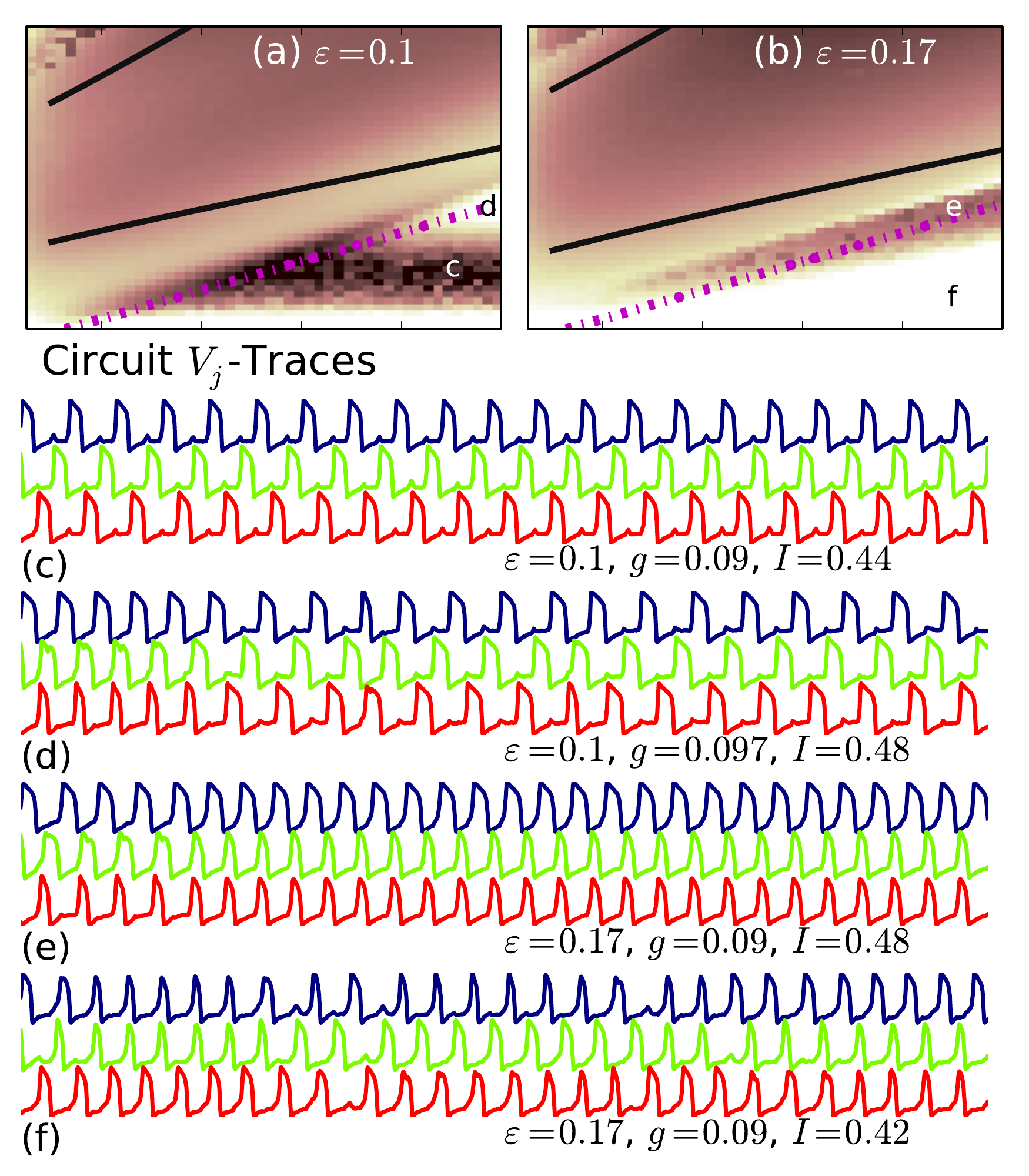}

\caption{(Color online) \textbf{Stochastic circuit dynamics at different levels
of robustness.} The examples illustrate specific regions of interest in the
complex rhythm robustness of stochastic circuit dynamics.  Colors (gray scales)
in Panels~(a) and~(b) are coded according to Fig.~\ref{fig:D_allfour}.}

\label{fig:D} \end{figure}

In the wave-rhythm regions in Fig.~\ref{fig:D_allfour}, the dependence of $D$
on parameters $I$, $g$, and $\eps$ is the most homogeneous.  Strengthening
inhibition, by increasing $g$, generally increases the local robustness of
polyrhythms against noise as explainable in the linear stability theory
(Sec.~\ref{sec:linear}).  However, we find a vastly complex behavior in the
region close to the release case (Fig.~\ref{fig:D}(a, b)).  At small $\eps$, a
strip of stable wave rhythmicity is observed (Fig.~\ref{fig:D}(c)).  By
shifting $I$ to larger values, the robustness of the circuit becomes less
pronounced.  In this region, linear theory predicts pacemaker rhythms to be
more stable (Fig.~\ref{fig:Floquet}(a)).  We speculate that increased stability
of pacemaker rhythms facilitate switching, because they can better serve as
intermediates in the switching process (Fig.~\ref{fig:D}(d)).

Generally, robust pacemaker rhythms can be achieved at larger values of $\eps$,
where the wave rhythms are less stable.  At $I=0.48$, $g=0.09$, and
$\eps=0.17$, for example, all five rhythms coexist, but the wave rhythm is
close to its stability boundary (cf.~Fig.~\ref{fig:bifDiag}(a)).  Here,
pacemaker rhythms were commonly observed in randomly perturbed traces of
circuit dynamics, as shown Fig.~\ref{fig:D}(e).  One might expect to enhance
stability of pacemaker rhythms by further reducing $I$ below the stability
boundary of the wave rhythms.  However, the circuit dynamics closer to the
release case turns out to be highly vulnerable, especially below the hard-lock
transition (dashed-dotted [pink] line).  In this region,  switching caused by
noise among the three coexisting pacemaker rhythms becomes very frequent
(cf.~Fig.~\ref{fig:D}(f)).  The increased intensity of switching is analogous
to the observations in three-cell motifs of the Hodgkin-Huxley type bursters in
Ref.~\cite{Schwabedal2014}, that demonstrated even weak noise can disrupt the
pacemaker rhythms subjected to the hard-lock inhibition occurring in the
release case.
The closer the circuit is set to the release case (lower $I$), the
higher the diffusion rate, $D$, becomes in the two-dimensional plane of phase
differences.

%None of these above effects can be properly explained by the Floquet exponents
%of individual circuit rhythms.  The stable (negative) exponent of pacemaker
%rhythms becomes increasingly small when $I$ is set to the release case
%(Fig.~\ref{fig:Floquet}(a)).  While this indicates more robust local dynamics,
%nevertheless the polyrhythms become increasingly vulnerable to noise.  The wave
%rhythms, on the other hand, become very unstable as they lose stability.  We
%hypothesize that this is also likely due to the hard-lock mechanism.   When
%hard locking is in effect, the phase relationships needed for the wave rhythms
%to occur become vulnerable to small perturbations.  The hard lock transition
%can therefore be viewed as an imperfect singular perturbation in the circuit. 

\subsection{Comparison of Methods for Stability Analysis}

We used several methods to carry out the qualitative and quantitative stability
analysis of the circuit polyrhythms.  The qualitative stability was assessed by
standard phase reduction, phase mappings, phase-basin analysis, and direct
bifurcation continuation.  The quantitative stability was assessed by local
methods of stability analysis derived from standard phase reduction, and the
Floquet exponents of continued periodic orbits.  It was also counterposed to
random perturbation analysis in its effect on phase diffusion.\\

\noindent\textbf{Qualitative Methods.}
Geometric and symmetry arguments in an all-to-all coupled circuit of
oscillators grants the existence of periodic orbits \cite{Golubitsky2006,
Antoneli2006, Antoneli2006, Dias2010, Golubitsky2012}.  However, these
arguments cannot be applied to show whether the corresponding rhythms (orbits)
are stable or not.  Methods of automatic bifurcation analysis can answer this
question successfully as demonstrated in this study
(cf.~Fig.~\ref{fig:bifDiag}).  The approach fails, however, if circuit rhythms
bifurcate.  In the example of the wave rhythm, a torus bifurcation leaves the
associated periodic orbit unstable;  disregarding the resulting low-amplitude
jiggle, the wave rhythm is still intact.  In the example, the torus-bifurcation
line (green dashed line in Fig.~\ref{fig:bifDiag}(a)) still coincides well with
the border of wave instability, but only by the coincidence that the torus is
stable for a small range of parameters.  In such cases, the methods of phase
description, such as return maps, are able to qualify that the torus orbit is
still \textit{close} to the wave rhythm (cf.~Fig.~\ref{fig:waveRhythm_bif}(c)).

The phase approach is applied with three different methods in this work.  In
the weak coupling approximation, the standard method of phase reduction
describes phase perturbations of individual periodic orbits by the phase
resetting curve (PRC) to linear order.  The method has many numerical
advantages.  The PRC is obtained by only regarding an individual oscillator;
subsequently it can be used to explore the phase dynamics of arbitrarily large
networks.  Moreover, a high precision can be reached because the method does
not require forward integration of the full circuit dynamics.  One can
therefore compute fixed points, as well as their eigenvalues of the phase
flow, as demonstrated in Fig.~\ref{fig:prcBifDiag}.

As coupling is strengthened, the standard phase reduction fails to produce
correct results because the individual periodic orbits become increasingly
distorted by the inhibitory coupling.  However, in principle, the phase
dynamics remain slow compared to that for the amplitudes.  Therefore, one can
still compute the return maps from first return times of individual
oscillations to reconstruct the phase flow.  The distance between consecutive
phase lags in the mapping will increase with stronger coupling because the
phase dynamics become fast compared to the oscillation period.  Eventually,
phase trajectories are not discernible anymore and the phase-mappings method
will fail.

Nevertheless, when phases jump erratically, it is still possible to
re-construct some practical aspects of the phase dynamics, for example, the
basins of attraction and their boundaries.  In the brute-force scheme, the
initial conditions can only form a topological equivalent of the individual
periodic orbit.  Therefore, the geometry of the basins is strongly distorted.
For example, the basins of the pacemaker rhythms may not appear equally sized,
even though they are in this symmetric circuit (cf.~Fig.~\ref{fig:torus}(d)).\\

\noindent\textbf{Quantitative Methods.}
Harder than the stability or instability of a circuit rhythm is the
evaluation how stable a rhythm is.  We used two approaches, in this work, to
quantify stability depending on the assumed disturbances to the circuit
dynamics:  approaches of linear stability measure the effect of infinitesimal
perturbations, whereas approaches with finite perturbations such as noise
investigate the full polyrhythmic stability, or robustness.

Infinitesimal perturbations cannot excite the circuit to switch from a stable
rhythm to another.  Therefore, the wave and pacemaker rhythms have to be
regarded separately.  From such an element-wise description of polyrhythmicity,
it is hard to predict the outcome of the switching behavior through Kramer's
rates, for example.  This would only be possible if one were able to quantify a
height of an assumed potential barrier separating stable circuit rhythms.
Phase diffusion coefficients of noise-perturbed circuit dynamics, on the other
hand, quantify random transitions between stable circuit rhythms.  This measure
is typically dominated by the most stable circuit rhythms, as highlighted in
the examples in Fig.~\ref{fig:D}.  The least stable rhythms take the role of
unstable saddles at finite noise strengths, but may also serve as facilitating
intermediates.

\section{Conclusions}

Small circuits of inhibitory relaxation oscillators appear in many natural
systems in order to flexibly generate rhythmic patterns of activity phases.  In
this article, we applied several computational methods to gain global
understanding of the dynamical transitions in a circuit of three mutually
inhibitory relaxation oscillators.  We find that the two wave and three
pacemaker rhythms, predicted to coexist in the circuit due to permutation
symmetry, strongly depend on quantitative and qualitative features of the node
dynamics and inhibitory coupling.

As a generic model of relaxation oscillations, we adopted a
Fitzhugh-Nagumo-like system that exhibits two saddle-node bifurcations,
beyond which oscillations stall.  One bifurcation inactivates the oscillators,
while the other stabilizes its active state.  These dynamical regimes are
non-generic for oscillators, but occur often in natural systems to facilitate
flexible control of frequency and rhythmicity.  Comparison of our results using
the generic model and those of \citet{Wojcik2011} and \citet{Jalil2010}
highlight to what extent the generic model can approximate Hodgkin-Huxley type
bursting models.  While the release case is well represented \cite{Wojcik2011},
the two models differ in the escape case where spikes play an active role in
rhythmogenesis \cite{Jalil2010}.

In our investigations, we find that closeness to bifurcations of individual
oscillators has a profound effect on the dynamics of the whole circuit:  a
generic inhibitory circuit produces the wave rhythms, but close to the
inactivating bifurcation, we find that the wave rhythms can become unstable to
give room for pacemaker rhythms.  We find the same dynamical instabilities in
the case of strong inhibition.  Past the hard-lock transition, all active
phases of mutually inhibitory neurons need to be non-overlapping.  This is
possible in our three-node circuit where, conversely, wave rhythms may still be
observed.  However, for circuits consisting of more nodes, waves can become
unstable due to such a crowding effect.

The method of phase basins described in this article gives a natural extension
to the return maps of \citet{Wojcik2011}, that allows for the treatment of
strong coupling in polyrhythmic circuits.  Tracking phase basins across
coupling strengths allows us to identify bifurcations that can be utilized for
the control of rhythms in the circuit.  Such coupling control may be of
particular experimental relevance in neuroscience, because inhibitory synaptic
strengths are easily modifiable by chemical agents.

Quantitative stability of polyrhythmicity is particularly hard to explore
because transitions can occur at any phase of a stable periodic orbit to
another.  We add noise to the dynamics to excite this potentially large number
of switching paths.  For weak noise, only the most probable paths are excited,
thus, revealing a skeleton of vulnerability in the full polyrhythmic dynamics.
The adoption of phase diffusion to quantify such stability features has
advantages over other possible methods, such as deviants of recurrences
\cite{Schwabedal2014}, or coarse-grained Markov chain descriptions
\cite{Pisarchik2014}.  The main advantage is the intrinsic invariance of the
phase diffusion constant \cite{Schwabedal2014b}, that allows for a reliable
estimation of complex features of the circuit dynamics.  To understand the
unfolding complexity of polyrhtymic switching, more refined techniques of
stochastic analysis will be necessary.

\acknowledgments

J.~S.~was funded by DFG grant SCHW~1685/1.  D.~K.~was funded by the GSU Brain
and Behaviors program and NSF~REU grant DMS-1009591, and the GSU honor program.
A.~S.~acknowledges the support from NSF DMS-1009591, RFFI~11-01-00001, RSF
grant 14-41-00044 at the Lobachevsky University of Nizhny Novgorod and the
grant in the agreement of August 27, 2013 N~02.B.49.21.0003 between the
Ministry of education and science of the Russian Federation and Lobachevsky
State University of Nizhni Novgorod (Sections~2-4), as well as NSF~BIO-DMS
grant IOS-1455527.  We thank the acting members of the NEURDS (Neuro Dynamical
Systems) lab at GSU: A.~Kelley, K.~Pusuluri, T.~Xing, and J.~Collens  for
helpful discussions.  We are grateful to NVIDIA Co.~for their generous donation
of a Tesla GPU card. We thank A.~Kelley, B.~Chung, B.~L\"unsmann, and E.~Latash
for her useful comments on the manuscript.

\bibliographystyle{apsrev4-1}

%\bibliography{manuscript}
%

\end{document}